\documentclass{iopart}

\pdfoutput=1

\usepackage{graphicx,epsfig,sidecap}
\usepackage{bm}

\catcode`@=11 
\renewcommand\tableofcontents{%
  \section*{\contentsname}%
  \@starttoc{toc}%
}
\catcode`@=12

\newcommand{\proj}[1]{\mbox{$|#1\rangle \!\langle #1 |$}}

\def\be{\begin{equation}}
\def\ee{\end{equation}}

\def\bea{\begin{eqnarray}}
\def\eea{\end{eqnarray}}

\def\Tr{{\rm Tr}\,}

\def\s{\sigma}
\def\a{\alpha}

\begin{document}

\title[Entanglement of 2 disjoint intervals in $c=1$ theories]
{Entanglement entropy of two disjoint intervals in $c=1$ theories}

\author{Vincenzo Alba$^1$, Luca Tagliacozzo$^2$, Pasquale Calabrese $^3$}
\address{
$^1$ Max Planck Institute for the Physics of Complex Systems, N\"othnitzer Str. 38, 01187 Dresden, Germany,\\
$^2$ School of Mathematics and Physics, The University of Queensland, 
Australia,\\  
ICFO, Insitut de Ciencias Fotonicas, 08860 Castelldefels (Barcelona) Spain} 
\address{$^3$ Dipartimento di Fisica dell'Universit\`a di Pisa and INFN,
             Pisa, Italy. }

\date{\today}

\begin{abstract}

We study the scaling of the R\'enyi entanglement entropy of two disjoint blocks of critical lattice models described by conformal
field theories with central charge $c=1$. 
We provide the analytic conformal field theory result for the second order R\'enyi entropy for a free boson compactified on 
an orbifold describing the scaling limit of the Ashkin-Teller (AT) model on the self-dual line.
We have checked this prediction in cluster Monte Carlo simulations of the classical two dimensional AT model. 
We have also performed extensive numerical simulations of the
anisotropic Heisenberg quantum spin-chain with tree-tensor network techniques
that allowed to obtain the reduced density matrices of disjoint blocks of the
spin-chain and to check the correctness of the predictions for R\'enyi and
entanglement  entropies from conformal field theory.
In order to match these predictions, we have extrapolated the numerical results by properly taking into account
the corrections induced by the finite length of the blocks to the leading scaling behavior.
\end{abstract}

\maketitle

\section{Introduction}

Let us imagine to divide the Hilbert space ${\cal H}$ of a given quantum system into two parts ${\cal H}_A$ and ${\cal H}_B$ such that 
 ${\cal H}={\cal H}_A\otimes {\cal H}_B$.
When the system is in a pure state $|\Psi\rangle$, the bipartite entanglement between 
A and its complement B, 
can be measured in terms of the R\'enyi entropies \cite{Renyi} 
\be
S_A^{(n)}=\frac1{1-n}\log{\rm Tr}\,\rho_A^n\,,
\label{renyidef}
\ee
where $\rho_A={\rm Tr}_B\,\rho$ is the reduced density matrix of
the subsystem A, and $\rho=|\Psi\rangle\langle\Psi|$ is the
density matrix of the whole system. 
The knowledge of $S_A^{(n)}$ as a function of $n$ identifies univocally the
full spectrum of non-zero eigenvalues of $\rho_A$ \cite{cl-08}, 
and provides complementary information about the
entanglement to the one obtained from the von Neumann entanglement
entropy  $S_A^{(1)}$. 
Furthermore, the scaling of  $S_A^{(n)}$ with the size of A in the ground-state
of a one-dimensional system is more suited than $S_A^{(1)}$ to
understand if a faithful representation of the state in term of a
matrix product state can be or cannot be obtained  with polynomial
resources in the length of the chain \cite{mps,cv-09}.


For a one-dimensional critical system whose scaling limit is described by a conformal
field theory (CFT), in the case when A is an interval of length $\ell$ 
embedded in an infinite system, the asymptotic large $\ell$ behavior of the quantities determining the R\'enyi
entropies is  \cite{Holzhey,cc-04,Vidal,cc-rev}
\begin{equation}\fl \label{Renyi:asymp}
\Tr\rho_A^{n}
\simeq c_n \left(\frac{\ell}{a}\right)^{c(n-1/n)/6}\,,\qquad \Rightarrow
S_A^{(n)}\simeq\frac{c}6 \left(1+\frac1n\right)\log \frac{\ell}a +c'_n\,,
\end{equation}
where $c$ is the central charge of the underlying CFT and $a$ the inverse of an ultraviolet cutoff (e.g. the lattice spacing).
The prefactors $c_n$ (and so the additive constants $c'_n$) are non universal 
constants (that however satisfy universal relations \cite{fcm-10}).

\begin{figure}[t]
\includegraphics[width=0.9\textwidth]{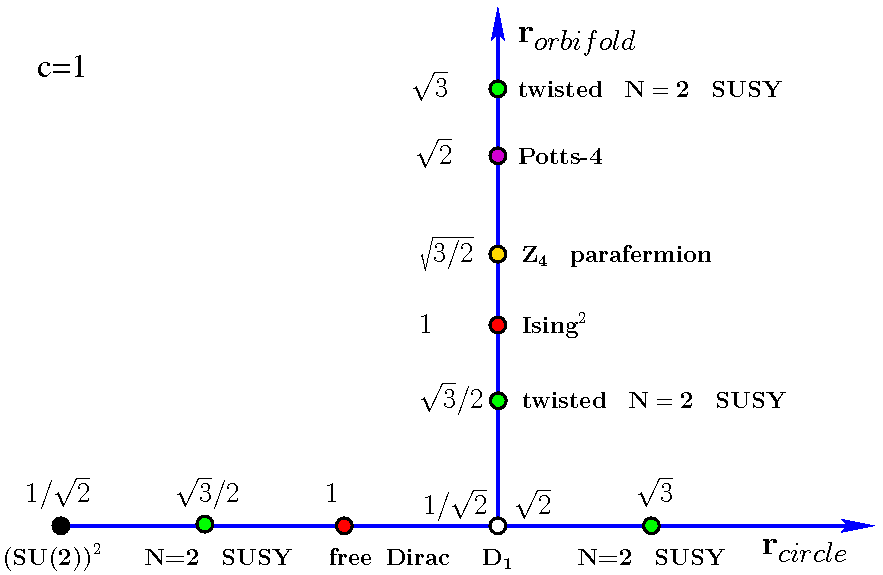}
\caption{Survey of  $c=1$ theories corresponding to a free boson
  compactified on a circle (horizontal axis) and on an orbifold (vertical axis) as reported e.g. in Refs. \cite{book}. 
  For some values of $r_{\rm circle}$ and $r_{\rm orb}$, the corresponding statistical mechanical models are reported.
  The XXZ spin chain in zero magnetic field lies on the horizontal axis in the interval $r_{\rm circle}\in[0,1/\sqrt2]$.
  The self-dual line of the Ashkin-Teller model lies on the vertical axis in the interval $ r_{\rm orb}\in[\sqrt{2/3},\sqrt{2}]$.}  
\label{fig14}
\end{figure}

The central charge is an ubiquitous and fundamental feature of a conformal field theory 
 \cite{c-lec}, but it does not always identify  the universality class of the theory.
A relevant class of relativistic massless quantum field theories are the $c=1$ models, which describe many
physical systems of experimental and theoretical interest. 
The one-dimensional Bose gas with repulsive interaction, the (anisotropic) Heisenberg spin chains, the Ashkin-Teller model 
and many others
are all described (in their gapless phases) by $c=1$ theories.
These are all  free-bosonic field theories where the boson field satisfies different periodicity constraints, i.e. it is
compactified on a specific target space. 
The two most notable examples are the compactification on a circle (corresponding to the Luttinger liquid field theory)
and on a $Z_2$ orbifold (corresponding to the Ashkin-Teller model \cite{z-87,book,dvv-87}). 
 The critical exponents 
 depend in a continuous way on the compactification radius of the bosonic field.
A survey of the CFTs compactified on a circle or on a $Z_2$ orbifold is given in Fig. \ref{fig14}, 
in a standard representation \cite{book,dvv-87}. 
The horizontal axis  is the compactification radius on the circle $r_{\rm circle}$, while
the vertical axis represents the value of the $Z_2$ orbifold compactification radius $r_{\rm orb}$. 
The two axes cross in a single point, meaning that the theories at  $r_{\rm circle}=\sqrt2$ and at $r_{\rm orb}=1/\sqrt2$
are the same. (The graph is not a cartesian plot, i.e. it has no meaning to have one $r_{\rm circle}$ and one $r_{\rm orb}$ at 
the same time.) 
For some values of $r_{\rm circle}$ and $r_{\rm orb}$, we report statistical mechanical models and/or field theories to 
which they correspond.
In the following we will consider the Ashkin-Teller model that on the self-dual line is described 
by  $ r_{\rm orb}\in[\sqrt{2/3},\sqrt{2}]$ and the XXZ spin chain in zero magnetic field that is described by   $ r_{\rm circle}\in[0,1/\sqrt2]$.
We mention that different compactifications  have been studied \cite{g-88}, but they correspond to more 
exotic statistical mechanical models and will not be considered here.

According to Eq. (\ref{Renyi:asymp}), the central charge of the CFT can be extracted
from the scaling of both the R\'enyi and von Neumann  entropies. 
In the last years, this idea  has overcome the previously available techniques of determining $c$, e.g.  by measuring the
finite size corrections to the ground state energy of a spin chain \cite{ecorr}. 
However, the dependence of the scaling of the  entropies of a single
block only on the central charge prevents to extract from them other
important parameter of the model such as the compactification radius.
%
It has been shown that instead  the entanglement entropies  of disjoint intervals are sensitive to the full operator content of the CFT 
and in particular they depend on the compactification radius and on the symmetries of the target space. 
Thus they encode complementary information about the underlying
conformal field theory of a given critical  quantum/statistical system
to the knowledge of the central charge present in the  scaling of the single block entropies.
(Oppositely in 2D systems with conformal invariant wave-function, the entanglement  entropy of a single region depends on the 
compactification radius \cite{2d}.)

This observation boosted an intense theoretical activity aimed at determining  R\`enyi entropies of disjoint intervals both 
analytically and numerically \cite{fps-08,cg-08,cct-09,ch-04,ffip-08,kl-08,rt-06,atc-09,ip-09,fc-10,h-10,fc-10b,c-10,cct-11}.
A part of  this paper is dedicated  to
consolidate some of the results already provided in other works where
they either have been studied only on very small chains, with the
impossibility of properly taking into account the severe finite size
corrections \cite{fps-08} or have been tested in the specific cases of spin
chains equivalent to free fermionic models \cite{atc-09,fc-10}.
An important point to recall when dealing with more than one
interval is that 
%
%
%
the R\'enyi entropies in Eq. (\ref{renyidef}) measure only the entanglement of the 
disjoint intervals with the rest of the system. 
They do {\it not}  measure the entanglement of one interval with respect to the other, that instead requires the 
definition of more complicated quantities because $A_1\cup A_2$ is in a mixed state (see e.g. Refs. \cite{Neg} for a 
discussion of this and examples). 
Furthermore, it must be mentioned that some results about the entanglement of two disjoint intervals are at the basis of 
a recent proposal to "measure" the entanglement entropy \cite{c-11}.

\subsection{Summary of some CFT results for the entanglement of two disjoint intervals}

We consider the case of two disjoint intervals 
$A=A_1\cup A_2=[u_1,v_1]\cup [u_2,v_2]$. 
By global conformal invariance, in the thermodynamic limit,  $\Tr \rho_A^n$ can be written as
\be\fl
\Tr \rho_A^n
=c_n^2 \left(\frac{|u_1-u_2||v_1-v_2|}{|u_1-v_1||u_2-v_2||u_1-v_2||u_2-v_1|} 
\right)^{\frac{c}6(n-1/n)} F_{n}(x)\,,
\label{Fn}
\ee 
where $x$ is the four-point ratio (for real $u_j$ and $v_j$, $x$ is real) 
\begin{equation}
x=\frac{(u_1-v_1)(u_2-v_2)}{(u_1-u_2)(v_1-v_2)}\,.
\label{4pR}
\end{equation}
The function $F_n(x)$ is a universal function (after being normalized such that $F_n(0)=1$) that encodes all the information about 
the operator spectrum of the CFT and in particular about the compactification radius. 
$c_n$ is the same non-universal constant  appearing in Eq. (\ref{Renyi:asymp}).

Furukawa,  Pasquier, and  Shiraishi \cite{fps-08} calculated
$F_2(x)$  for a free boson compactified on a circle of radius $r_{\rm circle}$
\be
F_{2}(x)=
\frac{\theta_3 (\eta \tau) \theta_3 (\tau/\eta )}{ [\theta_3 (\tau)]^{2}},
\label{F2}
\ee
where $\theta_\nu$ are Jacobi theta functions and the (pure-imaginary) $\tau$ is given by
\be
x= \left[\frac{\theta_2(\tau)}{\theta_3(\tau)}\right]^4\,,\qquad
\tau(x)=i\,\frac{ _2 F_1(1/2,1/2;1;1-x)}{ _2 F_1(1/2,1/2;1;x)}\,.
\label{mapping}
\ee 
$\eta$ is a universal critical exponent related to the compactification radius $\eta= 2r_{\rm circle}^2$. 
\footnote{Because of the symmetry $\eta\to1/\eta$ or $r_{\rm circle}\to 1/2 r_{\rm circle}$ for any conformal property 
one could also define $\eta= 1/2r_{\rm circle}^2$ as sometimes done in the literature.
However, corrections to scaling are not symmetric in $\eta\to1/\eta$ and this is often source of confusion. 
A lot of care should be used when referring to one or another notation.}
This has been extended to general integers $n\geq 2$ in Ref. \cite{cct-09}
\begin{equation}
F_n(x)=
\frac{\Theta\big(0|\eta\Gamma\big)\,\Theta\big(0|\Gamma/\eta\big)}{
[\Theta\big(0|\Gamma\big)]^2}\,,
\label{Fnv}
\end{equation}
where $\Gamma$ is an $(n-1)\times(n-1)$ matrix with elements \cite{cct-09}
\be
\Gamma_{rs} =  
\frac{2i}{n} \sum_{k\,=\,1}^{n-1} 
\sin\left(\pi\frac{k}{n}\right)\beta_{k/n}\cos\left[2\pi\frac{k}{n}(r-s)
\right]\,, 
\label{Gammadef}
\ee
and 
\be
\beta_y=\frac{\, _2 F_1(y,1-y;1;1-x)}{\, _2 F_1(y,1-y;1;x)}\,.
\label{betadef}
\ee
$\eta$ is the same as above, while $\Theta$ is the Riemann-Siegel theta function
\begin{equation}
\label{theta Riemann def}
\Theta(0|\Gamma)\,\equiv\,
\sum_{m \,\in\,\mathbf{Z}^{\a-1}}
\exp\big[\,i\pi\,m^{\rm t}\cdot \Gamma \cdot m\big]\,.
\end{equation}
The analytic continuation of Eq. (\ref{Fnv}) to real $n$ for general values of 
$\eta$ and $x$ (to obtain the von Neumann entanglement entropy) is still an open problem, but results for 
$x\ll1$ and $\eta\ll1$ are analytically known \cite{cct-09,cct-11}. 

The function $F_n(x)$ is known exactly for arbitrary integral $n$ also for the critical Ising field theory \cite{cct-11}. 
However, in the following we will need it only at $n=2$ (i.e. $F_2(x)$) for which it assumes the simple form \cite{atc-09}
\be\fl
F_2^{\rm Is}(x)=\frac1{\sqrt{2}}\Bigg[
\left(\frac{(1 + \sqrt{x}) (1 + \sqrt{1 - x})}2\right)^{1/2} + x^{1/4} + ((1 - x) x)^{1/4} + (1 - x)^{1/4} \Bigg]^{1/2}.
          \label{CFTF2}
\ee

In Ref. \cite{cct-11}, it has been proved that in any CFT the function $F_n(x)$ admits the small $x$ expansion
\be
F_n(x)=1+ \left(\frac{x}{4n^2}\right)^{ \a}s_2(n)+ \left(\frac{x}{4n^2}\right)^{2\a}s_4(n)+\dots \,,
\label{Fexpintro}
\ee
where $\a$ is the lowest scaling dimension of  the theory.
The functions $s_j(n)$ are calculable from a modification of the short-distance expansion \cite{cct-11}, and 
in particular it has been found \cite{cct-11} 
\be
s_2(n)={\cal N}\, \frac{n}2
\sum_{j=1}^{n-1}
\frac1{\left[ \sin\left(\pi\frac{j}{n}\right) \right]^{2\alpha}}\,,
\label{s2cft}
\ee
where the integer ${\cal N}$ counts the number of inequivalent correlation functions giving the same contribution.
This expansion  has been tested against the exact results for the free compactified boson (Ising model) 
with $\a=\min[\eta,1/\eta]$ ($\a=1/4$) and ${\cal N}=2$ (${\cal N}=1$).

All the results we reported so far are valid for an infinite system. 
Numerical simulations are instead performed for finite, but large, system sizes.
According to CFT \cite{cc-rev}, we obtain the correct result for a chain of finite length $L$  by
replacing all distances $u_{ij}$ with the {\it chord distance} $L/\pi \sin(\pi u_{ij}/L)$ (but different finite size forms exist 
for excited states \cite{abs-11}).
In particular the single interval entanglement is \cite{cc-04} 
\begin{equation} 
\label{SnFS}
\Tr\rho_A^{n}
\simeq c_n \left[\frac{L}{\pi a} \sin\left(\frac{\pi \ell}{L}\right)\right]^{-c(n-1/n)/6}\,,
\end{equation}
and for two intervals, in the case the two subsystems $A_{1}$ and $A_{2}$ have the same length $\ell$ and are placed at 
distance $r$, the four-point ratio $x$ is
\be
x=\left(\frac{\sin\pi\ell/L}{\sin\pi (\ell+r)/L}\right)^2\,.
\label{xFS}
\ee

\subsection{Organization of the paper}

In this paper we provide accurate numerical tests for the functions $F_n(x)$ in truly interacting lattice models 
described by a CFT with $c=1$.
In Sec.  \ref{sec2} we derive the CFT prediction for the function $F_2(x)$ of a free boson compactified on an orbifold
describing, among the other things, the self-dual line of the  AT model when $r_{\rm orb}\in[\sqrt{2/3},\sqrt{2}]$.
In order to check this result, we needed to develop a classical Monte Carlo algorithm in Sec. \ref{ATmc} based on the ideas
introduced in Ref. \cite{cg-08}. 
This algorithm is used in Sec. \ref{ATres} to determine $F_2(x)$  for several points on the self-dual line.    
We also consider the XXZ spin-chain in zero magnetic field to test the correctness of  Eq. (\ref{Fnv}).
In order to extend the results of Ref. \cite{fps-08} to longer
chains, we have used a tree tensor network algorithm that has allowed us
to study chains of length up to $L=128$ with periodic boundary conditions.
In this way, we have been able to perform a detailed finite size
analysis  that was difficult solely with the data from exact diagonalization reported in Ref. \cite{fps-08}. 
The analysis also shows that only through the knowledge of the unusual corrections to
the leading scaling behavior \cite{ccen-10,ce-10,cc-10,ccp-10,xa-11,fc-10} we are able to perform a  quantitative test of Eq. (\ref{Fnv}).
The tree tensor network algorithm is described in Sec. \ref{ttn:sec}, while the numerical results are presented in Sec. \ref{XXZ:sec}. 
The various sections  are independent one from each other, 
so that readers interested only in some results should have an easy access to them without reading the whole paper.  

\section{$n=2$ R\`enyi entanglement entropy for two intervals in the Ashkin-Teller model}
\label{sec2}

In a quantum field theory $\Tr\rho_A^n$ for integer $n$ is proportional to the partition function on an $n$-sheeted Riemann surface 
with branch cuts along the subsystem $A$, i.e. $\Tr\rho_A^n=Z_n(A)/Z_1^n$ where $Z_n(A)$
is the partition function  of the field theory on a conifold where $n$ copies of the manifold ${\cal R}={\rm system}\times R^1$
are coupled along branch cuts along each connected piece of $A$ at a time-slice $t=0$ \cite{cc-rev,cc-05p}.
Specializing to CFT, for a single interval on the infinite line, this equivalence leads to Eq. (\ref{Renyi:asymp}) \cite{cc-04}, 
whose analytic continuation to non-integer $n$ is straightforward.  
When the subsystem $A$ consists of $N$ disjoint intervals (always in an infinite system), 
the $n$-sheeted Riemann surface ${\cal R}_{n,N}$ has genus $(n-1)(N-1)$ and cannot be mapped to the complex plane so that 
the CFT calculations become more complicated. 

However, for two intervals ($N=2$), when for a given theory the partition function on a generic Riemann surface of genus
 $g$ with arbitrary {\it period matrix} is known,  $\Tr\rho_A^n$ can be easily deduced exploiting the results 
 of Refs. \cite{cct-09,cct-11}. 
In fact, a by-product of  the calculation for the free boson  \cite{cct-09} is that the $(n-1)\times (n-1)$ period matrix is always given  
by Eq. (\ref{Gammadef}).
Although derived for a free boson, the period matrix is a pure geometrical object and it is only related to the structure of the world-sheet 
${\cal R}_{n,2}$ and so it is the same for any theory.
This property has been used in Ref. \cite{cct-11} to obtain $F_n(x)$ for the Ising universality class for any $n$, in agreement with 
previously known numerical results \cite{fc-10}.
When also $n=2$, the surface ${\cal R}_{2,2}$ is topologically equivalent to a torus for which the partition function is known 
for most of the CFT. 
The torus modular parameter $\tau$ is related to the four-point ratio by Eq. (\ref{mapping}).
Thus, the function $F_2(x)$ is proportional to the torus partition function where $\tau$ is given by Eq. (\ref{mapping}) and with
the proportionality constant fixed by requiring $F_2(0)=1$.
This way of calculating $S^{(2)}_A$ is much easier than the general one for $S^{(n)}_A$ \cite{Dixon,cct-09} and indeed it has 
been used to obtain the first results both  for the free compactified boson \cite{fps-08} and for the Ising model \cite{atc-09}.

For a conformal free bosonic theory with action
\begin{eqnarray}
S=\frac{1}{2\pi}\int d z d \bar{z} \,\partial\phi\bar{\partial}\phi\,,
\label{fba}
\end{eqnarray}
the torus partition functions are known exactly both for circle and orbifold compactification \cite{tori,s-87,book}.

We now recall some well-known facts in order to fix the notations and derive the function $F_2(x)$ for the Ashkin-Teller
model.
The bosonic field $\phi$ is said to be compactified on a circle of radius $r_{\rm circle}$ when $\phi=\phi+2\pi r_{\rm circle}$.
The torus partition function (and the one on the $n$-sheeted Riemann surface) should be derived with this constraint. 
It is a standard CFT exercise to calculate the resulting torus partition function \cite{tori,book}
\be
Z_{\rm circle}(\eta)= \frac{\theta_3(\eta\tau) \theta_3(\tau/\eta)}{|\eta_D(\tau)|^2}\,,
\ee
where $\eta_D(\tau)$ is the Dedekind eta function and  $\eta=2r_{\rm circle}^2$.
Using Eq. (\ref{mapping}) and some properties of the elliptic functions, Eq. (\ref{F2}) for $F_2(x)$ follows  \cite{fps-08}. 
When specialized at $\eta=1/2$ (or $\eta=2$), $F_2(x)$ has the simple form
\be
F_2^{\rm XX}(x)= \sqrt{(1+x^{1/2})(1+(1-x)^{1/2})/2}\,,
\label{F2XX}
\ee
that describes the XX spin-chain (that is equivalent to free fermions via the non-local Jordan-Wigner transformation). 

The concept of orbifold emerges naturally in the context of theories
whose Hilbert space admits some discrete symmetries. 
Let us assume that $G$ is a discrete symmetry. 
For the free bosonic theory,  the simplest example is the one we are interested in, i.e.  the $Z_2$ symmetry. 
It acts on the point of the circle $S^1$ in the following way
\begin{eqnarray}
g:\phi\rightarrow -\phi\,.
\end{eqnarray}
For the partition function of  a theory on the torus, we introduce the notation \cite{book}
\be
\quad\begin{picture}(2,2)
\put(0,-4){\framebox(15,15)}
\put(3.5,-14){$\pm$}
\put(-9,1.5){$\pm$}
\end{picture}\quad\quad
\ee
\vspace{3mm}
 where the $\pm$ denotes the boundary conditions on the two directions on
 the torus. The full partition function, given a finite discrete group $G$, is 
\begin{eqnarray}
Z_{{\cal T}/G}=\frac{1}{|G|}\sum\limits_{g,h\in G}\quad
\begin{picture}(2,2)
\put(0,-4){\framebox(15,15)}
\put(3.5,-14){\it h}
\put(-9,1.5){\it g}
\end{picture}
\label{modout}
\end{eqnarray} 
where $|G|$ denotes the number of elements in the group. 
The generalization to higher genus Riemann surfaces is straightforward (but it is not so easy  to obtain results, 
see e.g.  \cite{dvv-87,orb2}). 

Now we specialize Eq. (\ref{modout}) to the case of the $Z_2$ symmetry. 
Since the action (\ref{fba}) is invariant under $g:\phi\rightarrow -\phi$, 
we have the torus partition function for the free boson on the orbifold \cite{tori,s-87,book}
\be
Z_{orb}=\frac{1}{2}\bigg(\quad
\begin{picture}(2,2)
\put(0,-4){\framebox(15,15)}
\put(3.5,-14){$+$}
\put(-9,1.5){$+$}
\end{picture}\qquad +
\quad\begin{picture}(2,2)
\put(0,-4){\framebox(15,15)}
\put(3.5,-14){$+$}
\put(-9,1.5){$-$}
\end{picture}\quad\quad+
\quad\begin{picture}(2,2)
\put(0,-4){\framebox(15,15)}
\put(3.5,-14){$-$}
\put(-9,1.5){$+$}
\end{picture}\quad\quad+
\quad\begin{picture}(2,2)
\put(0,-4){\framebox(15,15)}
\put(3.5,-14){$-$}
\put(-9,1.5){$-$}
\end{picture}\quad\quad
\bigg)\,.
\ee
\vspace{1mm}
Standard CFT calculations lead to the result \cite{book}
\begin{eqnarray}
Z_{\rm orb}(\eta)=\frac{1}{2}\bigg(Z_{\rm circle}(\eta)+\frac{|\theta_3\theta_4|}{\eta_D\bar{\eta}_D}+\frac{|\theta_2\theta_3|}{\eta_D\bar{\eta}_D}+ 
  \frac{|\theta_2\theta_4|}{\eta_D\bar{\eta}_D}\bigg)\,,
\end{eqnarray}
where all the $\tau$ arguments in $\theta_\nu$ and $\eta_D$  are understood.
At the special point $\eta=1/2$ (or $\eta=2$) we get
\begin{eqnarray}\fl
Z_{\rm orb}(\eta=1/2)=\frac{1}{2}\bigg(\frac{|\theta_3|^2+|\theta_4|^2+|\theta_2|^2}{2|\eta_D|^2}+
  \frac{|\theta_3\theta_4|}{\eta_D\bar{\eta}_D}+\frac{|\theta_2\theta_3|}{\eta_D\bar{\eta}_D}+ 
  \frac{|\theta_2\theta_4|}{\eta_D\bar{\eta}_D}\bigg)={Z}_{\rm Ising}^2\,.
\label{atpft}
\end{eqnarray}

\begin{figure}[t]
\begin{center}
\includegraphics[width=.8\textwidth]{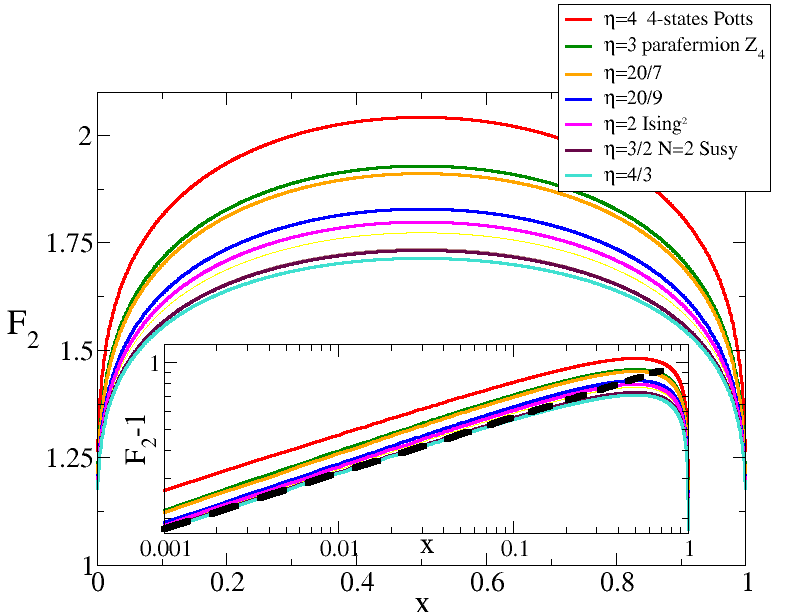}
\end{center}
\caption{$F_2(x)$ for the Ashkin-Teller model on the self-dual line for some values of $\eta$. 
Inset:  $F_2(x)-1$ in log-log scale to highlight the small $x$ behavior. The black-dashed line is $\sim x^{1/4}$. }
\label{log_curve}
\end{figure}

Thus, from the orbifold partition function, using the last identity and normalizing such that $F_2^{AT}(0)=1$, 
we can write the funcion $F_2^{\rm AT}(x)$ as
\be
F_2^{\rm AT}(x)=\frac{1}{2}\bigg(F_2(x)-F_2^{XX}(x)\bigg)+(F_2^{\rm Is}(x))^2\,,
\label{atf2}
\ee
where $F_2(x)$ is given in Eq. (\ref{F2}),
$F_2^{XX}(x)$ is the same at $\eta=1/2$ (cf. Eq. (\ref{F2XX})) and $F_2^{\rm Is}(x)$ is the result for
Ising (cf. Eq. (\ref{CFTF2})). 
As a consequence of the $\eta\leftrightarrow 1/\eta$ symmetry of $F_2(x)$, also $F_2^{AT}(x)$ displays the same invariance.
For small $x$, recalling that $F_2(x)- 1\sim x^{{\rm min}[\eta,\eta^{-1}]}$,
$F_2^{XX}- 1\sim x^{1/2}$ and $F_2^{\rm Is}- 1\sim x^{1/4}$, we have 
\be
 F_2^{AT}(x)-1\sim 
 \cases{
 x^{1/4}& for $\eta\ge 1/4$\,,\\
x^{{\rm min}[\eta,\eta^{-1}]}  & for $\eta\le 1/4$\,.
}
\ee
The critical Ashkin-Teller model lies in the interval $\sqrt{2/3}<r_{\rm orb}<\sqrt{2}$ and so
$4/3<\eta=2r_{\rm orb}^2< 4$. 
Thus we have $F_2^{AT}(x)-1\sim x^{1/4}$ along the whole self-dual line.
$F_2^{\rm AT}(x)$  for various values of $\eta$ in the allowed range is reported in Fig. \ref{log_curve}, 
where the behavior for small $x$ is highlighted in the inset to show the constant $1/4$ exponent.

\section{The classical Ashkin-Teller model and the Monte Carlo simulation}
\label{ATmc}

\begin{figure}[t]
\begin{center}
\includegraphics[width=.9\textwidth]{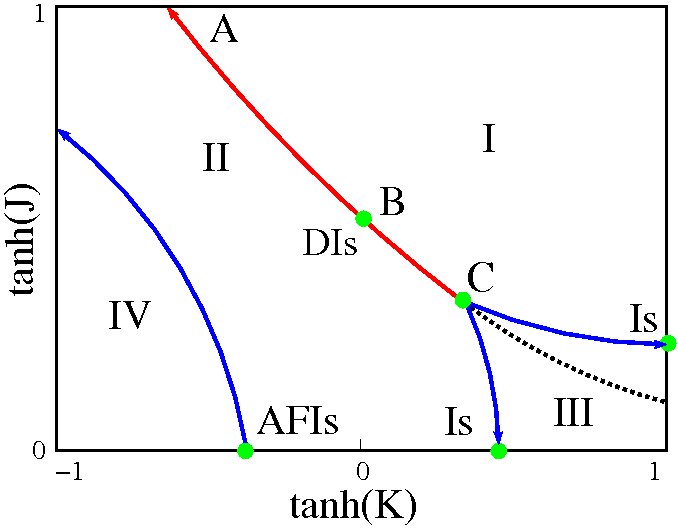}
\end{center}
\caption{Phase diagram of the 2D symmetric Ashkin-Teller model defined by the Hamiltonian (\ref{sat}). 
The red ABC line is the self dual line. 
The point $B$ at $K=0$ corresponds to two uncoupled Ising models. 
The point $C$ is the critical four-state Potts model at $K=J=(\log 3)/4$. 
At $J=0$ there are two critical Ising points at $K=\pm(\log(1+\sqrt{2}))/2$,  one (Is)  ferromagnetic and
the other (AFIs) antiferromagnetic. 
For $K\rightarrow\infty$ there is another critical Ising point at $J=(\log(1+\sqrt{2}))/2$. 
All continuous lines are critical. 
The blue lines $C-Is$ and the one starting at $AFIs$ are in the Ising universality class. 
The red line is critical with continuously varying critical exponents. 
The region denoted by I corresponds to a ferromagnetic phase for all the variables.
In the region II, $\s$, $\tau$, and $\s\tau$ are paramagnetic. 
In the region III only $\sigma\tau$ is ferromagnetic and in region IV $\sigma\tau$ exhibits
antiferromagnetic order while $\sigma$ and $\tau$ are paramagnetic.} 
\label{phadia}
\end{figure}

The two dimensional Ashkin-Teller (AT) model on a square
lattice is defined by the
Hamiltonian
\begin{eqnarray}
H=J\sum\limits_{\langle ij\rangle}\sigma_i\sigma_j +
J'\sum\limits_{\langle ij\rangle}\tau_i\tau_j+K\sum
\limits_{\langle ij\rangle}\sigma_i\sigma_j\tau_i\tau_j\,,
\label{ash}
\end{eqnarray}
where $\sigma_i$ and $\tau_i$ are classical Ising variables (i.e. can assume only the values $\pm1$). 
Also the product $\sigma\tau$ can be considered as an Ising variable. 
The model has a rich  phase diagram whose features are reported in full details in Baxter's book \cite{BB}.
We review in the following only the main features of this phase diagram. 
Under any permutation of the variables $\sigma,\tau,\sigma\tau$  the
AT model is mapped onto itself. 
At the level of the coupling constants, this implies that the model is invariant under any
permutation of $J,J',K$.  
For $K=0$, the AT model corresponds to two decoupled Ising models in 
$\sigma$ and $\tau$ variables. 
For $K\rightarrow \infty$ it reduces to a single Ising model with coupling constant $J+J'$.
For  $J=J'=K$ it corresponds to the four-state Potts model. 
It is useful to restrict to the symmetric Ashkin-Teller model where $J=J'$
\begin{eqnarray}
H=J\sum\limits_{\langle ij\rangle}(\sigma_i\sigma_j +
\tau_i\tau_j)+K\sum
\limits_{\langle ij\rangle}\sigma_i\sigma_j\tau_i\tau_j\,.
\label{sat}
\end{eqnarray}
The full phase diagram is reported in Fig. \ref{phadia} (in units of the inverse temperature $\beta=1$).
The model corresponds to two decoupled critical Ising models at $K=0$
and $2J=\log(1+\sqrt{2})$. For $J=0$ it is equivalent to
a critical Ising model in the variable $\sigma\tau$ with critical points at
$2K_\pm=\pm\log(1+\sqrt{2})$.  
For $K\rightarrow\infty$ there are two critical Ising points at 
$2J=\pm\log(1+\sqrt{2})$. 
On the diagonal $J=K$ the system
corresponds to a 4-state Potts model which is critical at $K=(\log 3)/4$. 
The different kinds of orders appearing in the phase diagram are explained in the caption of  Fig. \ref{phadia}. 
All the continuous lines in Fig.~\ref{phadia} are {\it critical lines}. 
The blue lines C-Is are in the Ising universality class. The line starting from AFIs belongs to the
antiferromagnetic Ising universality class. 
On the red line ABC the system is critical and the critical exponents vary continuously \cite{cont,BB}.

The AT model on a planar graph can be mapped to another AT model on the dual graph. 
When specialized to the square lattice, the phase diagram is equivalent to its dual on the self-dual line:  
\begin{eqnarray}
e^{-2K}=\sinh(2J)\,.
\end{eqnarray}
On this line, the symmetric AT model maps onto an homogeneous
six-vertex model which is exactly solvable \cite{BB}. 
It follows that on the self-dual line the model is critical for $K\le(\log3)/4$ 
and its critical behavior is described by a CFT with $c=1$. 
Along the self-dual line the critical exponents vary countinuously and are exactly known. 
For later convenience it is useful to parametrize the self dual line by a new parameter $\Delta$ 
\be
e^{4J}=\frac{\sqrt{2-2\Delta}+1}{\sqrt{2-2\Delta}-1}\,,\qquad\qquad
e^{4K}=1-2\Delta\,,
\ee
with $-1<\Delta<1/2$. In terms of $\Delta$, the orbifold compactification radius is \cite{s-87}
\be
\eta={2r_{\rm orb}^2}=\frac{4\arccos (-\Delta)}\pi=\frac{2}{K_L}\,,
\label{etaAT}
\ee
where $K_L$ is the equivalent of the Luttinger liquid parameter for the AT model.

\subsection{Cluster representation and Monte Carlo simulation}

A Swendsen-Wang type cluster algorithm for the AT model  has been proposed in Ref. \cite{dom} and then
re-derived in a simpler way by Salas and Sokal \cite{ss}. 
Here we partly follow the derivation of Salas and Sokal and we restrict to the symmetric
AT Hamiltonian (\ref{sat}) and assume $J\ge|K|$.
Using the identities for Ising type variables 
\be
\sigma_i\sigma_j=2\delta_{\sigma_i\,\sigma_j}-1\,, \qquad
\tau_i\tau_j=2\delta_{\tau_i\,\tau_j}-1\,,
\ee
we can rewrite Eq. (\ref{sat}) as 
\be\fl
-H=J\sum\limits_{\langle ij\rangle}(2\delta_{\sigma_i\,\sigma_j}+
 2\delta_{\tau_i\,\tau_j}-2)+K\sum\limits_{\langle
 ij\rangle}(2\delta_{\sigma_i\,\sigma_j}-1)(2\delta_{\tau_i\,\tau_j}-1)\,.
\ee
For convenience we shift the interaction (\ref{sat}) by $-4J$. 
In order to write the Boltzmann weight associated to a specific
configuration we  use 
$\exp(w\delta_{\sigma_i\,\sigma_j})=(\exp(w)-1)\delta_{\sigma_i\,\sigma_j}+1$  
and the analogous identity for the $\tau$ variables. 
The Boltzmann weight of a given link $\langle ij\rangle$ is then
\bea
{\cal{W}}_{\langle  ij\rangle}(\sigma_i,\sigma_j,\tau_i,\tau_j)&=&e^{-4J}+[e^{-2(J+K)}-e^{-4J}][\delta_{\sigma_i
 \,\sigma_j}+\delta_{\tau_i\,\tau_j}]+\nonumber \\ &&+
[1-2e^{-2(J+K)}+e^{-4J}]\delta_{\sigma_i\,\sigma_j}\delta_{\tau_i\,\tau_j}\,.
\label{atw}
\eea

The key idea for the Swendsen-Wang algorithm is to introduce two new auxiliary Ising-type variables $m_{ij}$ and $n_{ij}$ 
living on the link $\langle ij\rangle$. 
We redefine the Boltzmann weight on the link
$\langle ij\rangle$ as \cite{ss}
\bea\fl
{\cal{W}}_{\langle ij\rangle}&&(\sigma_i,\sigma_j,\tau_i,\tau_j,m_{ij},n_{ij})=
  e^{-4J}\delta_{m_{ij} 0}\delta_{n_{ij}0}+ \nonumber \\\fl&&
   +[e^{-2(J+K)}-e^{-4J}][  \delta_{\sigma_i \sigma_j}\delta_{m_{ij}1}\delta_{n_{ij}0}+\delta_{\tau_i\tau_j}
  \delta_{m_{ij}0}\delta_{n_{ij}1}]+ \nonumber\\ \fl&&+
[1-2e^{-2(J+K)}+e^{-4J}]\delta_{\sigma_i \sigma_j}\delta_{\tau_i \tau_j}
  \delta_{m_{ij}1}\delta_{n_{ij}1} \,.
\label{atwc}
\eea
Summing over $m_{ij}$ and $n_{ij}$ we obtain the weight in Eq. (\ref{atw}). 
Eq. (\ref{atwc}) has a graphical interpretation in terms of   clusters. 
In fact we can divide the links of the lattice in ``activated'' (if $m_{ij}=1$) or ``inactive'' (if $m_{ij}=0$). 
The same considerations hold for the $n_{ij}$ variables. 
Therefore, each link of the lattice can be activated by setting $m_{ij}=1$ or $n_{ij}=1$. 
The active links connect different lattice sites forming  clusters. 
There are clusters referring to the $\sigma$ variables (called $\sigma$-clusters) and to the $\tau$ variables ($\tau$-clusters). 
Isolated lattice sites are clusters as well. 
Obviously, the lattice sites belonging to the $\sigma$-clusters  ($\tau$-clusters) have the same value of $\sigma$ ($\tau$). 
The partition function of the extended model defined by the weight (\ref{atwc}) can be written as
\be
Z=\sum\limits_{\sigma,\tau=\pm 1}\sum\limits_{m,n=\pm 1}\prod\limits_{\langle ij\rangle}{\cal W}_{\langle
  ij\rangle}(\sigma_i,\sigma_j,\tau_i,\tau_j,m_{ij},n_{ij})\,. 
\ee
We now proceed to the following definitions. 
We divide all the links into three classes: we define $l_0$ the total 
number of inactivated links; $l_1$ the total number of links
connecting sites which belong only to one type of clusters either a
$\sigma$-cluster or a $\tau$-cluster.  We define $l_2$ the total
number of links on which $m$ and $n$ are both equal to $1$. 
Furthermore we introduce the quantities 
\begin{eqnarray}
B_0\equiv e^{-4J}\,,\\
B_1\equiv [e^{-2(J+K)}-e^{-4J}]\,,\\
B_2\equiv [1-2e^{-2(J+K)}+e^{-4J}]\,.
\end{eqnarray}
The following step is to perform  the summation over $\sigma,\tau$ in Eq. (\ref{atwc}). 
This is readily done,  obtaining the final expression for the partition function
\begin{eqnarray}
Z=\sum\limits_{{\cal C}\{\tau,\sigma\}}B_0^{l_0}B_1^{l_1}B_2^{l_2}\,2^{C^{\sigma}+C^{\tau}}  \,,
\label{atpf}
\end{eqnarray}
where we denoted with $C^{\sigma}$ the number of $\sigma$-clusters and
with $C^\tau$ the total number of $\tau$-clusters. 
In the counting of  $\tau$-clusters ($\sigma$-clusters) we included all the lattice
sites connected by a link on which $m_{ij}=1$ ($n_{ij}=1$). 
Isolated sites (with respect to $m$ or $n$ or both) count as single clusters. 
The links where $m_{ij}=1,n_{ij}=1$ contribute to both types of clusters.

\begin{figure}[t]
\begin{center}
\includegraphics[width=0.5\textwidth]{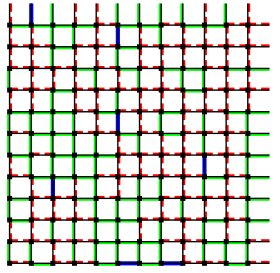}
\end{center}
\caption{A typical cluster configuration on a $12\times 12$ lattice. 
 Green lines are $\sigma$-clusters and red dashed lines are $\tau$-clusters. 
 Links in blue are double links.  Periodic boundary conditions on both directions are used.} 
\label{clu_ins}
\end{figure}

\subsection{Swendsen-Wang algorithm (the direct and embedded algorithms)}

We are now in position to write the Swendsen-Wang algorithm for the symmetric AT model. 
The Monte-Carlo procedure can be divided in two steps. 
In the first one, given a configuration for $(\sigma,\tau)$ variables, we construct a configuration of the
$(m,n)$ variables. In the second step we update the $(\sigma,\tau)$
variables at given $(m,n)$. 
The details of the step one are
\begin{itemize}
\item if $\sigma_i=\sigma_j$ and $\tau_i=\tau_j$, we choose
  $(m_{ij},n_{ij})$ with the following probabilities:
\begin{itemize}
\item $(m_{ij},n_{ij})=(1,1)$ with $p_1=1-2e^{-2(J+K)}+e^{-4J}$,
\item $(m_{ij},n_{ij})=(1,0)$ with $p_2=e^{-2(J+K)}+e^{-4J}$,
\item $(m_{ij},n_{ij})=(0,1)$ with $p_2=e^{-2(J+K)}+e^{-4J}$,
\item $(m_{ij},n_{ij})=(0,0)$ with $p_3=1-p_1-2p_2$,
\end{itemize}
\item if $\sigma_i=\sigma_j$ and $\tau_i=-\tau_j$, the probabilities are
\begin{itemize}
\item $(m_{ij},n_{ij})=(1,0)$ with $p_1=1-e^{-2(J-K)}$,
\item $(m_{ij},n_{ij})=(0,0)$ with $p_2=1-p_1$,
\end{itemize}
\item if $\sigma_i=-\sigma_j$ and $\tau_i=\tau_j$, the probabilities are
\begin{itemize}
\item $(m_{ij},n_{ij})=(1,0)$ with $p_1=1-e^{-2(J-K)}$,
\item $(m_{ij},n_{ij})=(0,0)$ with $p_2=1-p_1$,
\end{itemize}
\item if $\sigma_i=-\sigma_j$ and $\tau_i=-\tau_j$ we choose
  $(m_{ij},n_{ij})=(0,0)$ with probability $1$.
\end{itemize} 
In the step two, given the configuration of $(m,n)$ generated using the
rules above we build the connected $\sigma$-clusters and 
$\tau$-clusters. The value of $\sigma$ ($\tau$) spins are
required to be equal within each $\sigma$-cluster ($\tau$-cluster). 
We choose randomly the spin value in each cluster and independently of
the value assumed on the other clusters. This completes the update scheme. 
(Note a typo in Ref. \cite{ss}: the minus sign in step 2 and 3 of the update is missing.)

In Ref. \cite{ss} also the so called embedded version of the cluster algorithm is introduced.
Its implementation is slightly easier compared to the direct algorithm. 
In the embedded algorithm instead of treating both $\sigma$ and $\tau$ at the same time, one
deals with only one variable per time. 
Let us consider the Boltzmann weight of a link $\langle ij\rangle$ at fixed configuration of $\tau$
\begin{eqnarray}
{\cal W}_{\langle
  ij\rangle}(\sigma_i,\sigma_j,\tau_i,\tau_j)=e^{-2(J+K\tau_i\tau_j)}+
  (1-e^{-2(J+K\tau_i\tau_j)})\delta_{\sigma_i\,\sigma_j} \,.
\end{eqnarray}   
The model defined by this weight can be simulated with a standard Swendsen-Wang algorithm for the Ising model
 using  the effective coupling constant
\begin{eqnarray}
J^{eff}_{ij}=J+K\tau_i\tau_j\,.
\end{eqnarray}
This is no longer translation invariant, but this does not affect the effectiveness of the cluster algorithm for the Ising model as long as  
$J^{eff}_{ij}\ge 0$. 
The same reasoning applies to the case of fixed $\sigma$. 
Thus, the embedded algorithm is made of two steps
\begin{itemize}
\item For a given configuration of $\tau$ variables, we apply a standard
  Swendsen-Wang algorithm to $\sigma$ spins. The probability arising in 
  the update step is $p_{ij}=1-e^{-2(J+K\tau_i\tau_j)}$.
\item For a given configuration of  $\sigma$ variables, we update $\tau$ with
  the same algorithm and  probability  $p_{ij}=1-e^{-2(J+K\sigma_j\sigma_i)}$.  
\end{itemize}
Direct and embedded algorithms are both extremely effective 
procedures to sample the AT configurations. 
However,  very important for the following, Eq. (\ref{atpf}) for the
partition function does not hold anymore for a $n$-sheeted Riemann surface and we do not know whether it is possible 
to write the embedded algorithm for this case.

\subsection{R\'enyi entanglement entropies via Monte Carlo simulation of a classical system.}
In this section we summarize the method introduced by Caraglio and
Gliozzi \cite{cg-08} to obtain the R\'enyi entropies via simulations of classical systems and we generalize it to the AT model. 
The partition function $Z=\Tr e^{-\beta H}$ of a $d$-dimensional quantum system at
inverse temperature $\beta$ can be written as an Euclidean path integral in $d+1$ dimensions \cite{cc-rev}.
Thus for the $n$-th power of the  partition function one has 
\begin{eqnarray}
Z^n=\int\prod\limits_{k=1}^n{\cal D}[\phi_k]e^{-\sum\limits_{k=1}^n S(\phi_k)}
\end{eqnarray}
where $\phi_k\equiv\phi_k(\vec x, \tau)$ is a field living on the $k$-th replica of the system and $S(\phi_k)$ is the euclidean
action ($\tau$ is the imaginary time.) 
The actual form of the action is not important, but for the
sake of simplicity we restrict to the case of  nearest-neighbor interactions 
\begin{eqnarray}
S(\phi_k)=\sum\limits_{\langle ij\rangle}F(\phi_k(i),\phi_k(j))\,,
\end{eqnarray} 
and the function $F$ is arbitrary. 
We recall that $\Tr\rho_A^n$ can be obtained by considering the euclidean partition function over 
a $n$-sheeted Riemann surface with branch cuts along the subsystem $A$ \cite{cc-rev}. 
(This equivalence is also the basis of all quantum Monte Carlo methods to simulate the block 
entanglement in any dimension \cite{qmc}.)
Caraglio and Gliozzi constructed this $n$-sheeted Riemann surface for the lattice model in the following way.
Let us consider a square lattice (for simplicity) and take the two points of its dual lattice surrounding $A$ (that in 1+1 dimension is just 
an interval with two end-points). 
The straight line joining them defines the cut that we call $\lambda$. 
The length of $\lambda$ is equal to the length of $A$. 
Let us consider $n$ independent copies of this lattice with a cut. 
The  $n$-sheeted Riemann lattice is defined by assuming that  all the links of the $k$-th replica  intersecting the cut  
connect with the next replica $k+1(\textrm{mod}\,n)$. 
To get the partition function over the $n$ sheeted  Riemann surface we define the
corresponding coupled action 
\be\fl
S^n(\phi_k)=\sum\limits_{k=1}^n\sum\limits_{\langle ij\rangle\notin
  \lambda}F(\phi_k(i),\phi_k(j))+ \sum\limits_{\langle ij\rangle\in
  \lambda}F(\phi_k(i),\phi_{k+1 ({\rm mod}\,n)}(j)) \,.
\label{act}
\ee
This definition can be used in any dimension, even though we will use here  only $d=2$.
Finally, calling $Z_n(A)$ the partition function over the action (\ref{act}),  $\Tr\rho^n_A$ is given by 
\begin{eqnarray}
\Tr\rho^n_A=\frac{Z_n(A)}{Z^n}\,.
\end{eqnarray}

Following Ref. \cite{cg-08} we  introduce the  observable
\begin{eqnarray}
{\cal O}\equiv
  e^{-S^n(\phi_1,\phi_2,\dots,\phi_n;\lambda)+\sum_{k=1}^n
  S(\phi_k;\lambda)} \,,
\label{obs}
\end{eqnarray}
where $S^n$ and $S$ are the euclidean actions of the model  defined on the $n$-sheeted lattice and on the $n$
independent lattices respectively. 
The sum is restricted to links crossing the cut, as the presence of $\lambda$ in the arguments stresses.   
It then follows
\be
\langle{\cal O}\rangle_n\equiv\frac{Z_n(A)}{Z^n}=\Tr\rho^n_A\,,
\ee
where $\langle\cdot\rangle_n$ stands for the average taken 
onto the uncoupled action $\sum_{k=1}^n S(\phi_k)$.

We can now discuss our improvement to the procedure highlighted so far. 
The practical implementation of Eq. (\ref{obs})  to calculate $\Tr\rho_A^n$ is
plagued by severe limitations: 
analyzing the Monte-Carlo evolution of the observable, one notices that it
shows a huge variance because it is defined by an exponential. 
Direct application of Eq. (\ref{obs}) is possible then only for small lengths of the subsystem $A$. 
In order to overcome this problem,  let us consider the quantity ${Z_n(A)}/{Z^n}$ and imagine to divide the
subsystem in $L$ parts to have $A=A_1\cup A_2\dots \cup A_L$, with the
lengths of the various parts being arbitrary. 
Moreover we define a set of  subsystems $\hat{A}_i\equiv \cup_{k=1}^i A_i$. 
Then it  holds 
\begin{eqnarray}
\frac{Z_n(A)}{Z^n}=\prod\limits_{i=0}^L\frac{Z_n(\hat{A}_{i+1})}{Z_n(\hat{A}_i)}\,.
\label{trick}
\end{eqnarray}
Eq. (\ref{trick}) is very useful because each term in the product can be simulated
effectively using a modified version of (\ref{obs}) if we choose
the length of $A_i$ to be small enough. 
In fact, by definition,  we have  
\begin{eqnarray}
\langle{\cal O}(\hat{A}_i)\rangle_{{\cal
    R}_n(\hat{A}_i)}\equiv\frac{Z_n(\hat{A}_{i+1})}{{\cal
    Z}_n(\hat{A}_i)} \,,
\label{trick1}
\end{eqnarray}
where ${\cal O}(\hat{A}_i)$ is the modified observable
\begin{eqnarray}
{\cal O}(\hat{A}_i)\equiv\exp(-S^n(\hat{A}_{i+1})+S^n(\hat{A_i}))\,.
\end{eqnarray}
We stress that in Eq. (\ref{trick1}) the expectation value in the l.h.s
must be taken on the coupled action on the Riemann surface
with cut $\hat{A}_i$. 
The disadvantage of Eq. (\ref{trick}) is that, to simulate  large subsystems, one has to perform $L$
independent simulations and then build the observable taking the product of the results. 
If the dimension of each piece $A_i$ is small this task requires a large computational effort. 
Another important aspect is the estimation of the Monte Carlo error: 
if each term in (\ref{trick}) is obtained independently,  the error in the product is
\begin{eqnarray}
\frac{\sigma({\cal O})}{\overline{\cal O}}=\sqrt{\sum\limits_{i=0}^L\frac{\sigma^2({\cal O}({\hat A}_i))}{{\overline{{\cal O}({\hat A}_i)}}^2}}\,.
\label{error}
\end{eqnarray} 
If  the lengths of the intervals $A_i$ are all equal, then the
single terms of the summation in Eq. (\ref{error}) do not change 
much and the total error should scale as $\sqrt{L}$. 

Caraglio and Gliozzi \cite{cg-08} used another strategy to circumvent the problem with the observable in Eq.  (\ref{obs}). 
The trick was to consider the Fortuin-Kastelayn cluster expansion of the partition function of the Ising model. 
The analogous for the AT model was reported in the previous section 
\begin{eqnarray}
Z=\sum\limits_{{\cal
    C}\{\sigma,\tau\}}B_0^{l_0}B_1^{l_1}B_2^{l_2}\,2^{C^{\sigma}+C^{\tau}} \,, 
\end{eqnarray}
where ${\cal C}^{\sigma,\tau}$ are the $\sigma/\tau$-cluster configurations. 
Going from $n$ independent sheets to the $n$-sheeted lattice, the type of
 links and their total number  do not change, but
 the number of clusters does change, and so we get the cluster  expression of observable (\ref{obs}) for the AT model 
\begin{eqnarray}
{\cal O}(\hat{A_i})= 2^{[C_\sigma(\hat{A}_{i+1})+C_\tau(\hat{A}_{i+1})
 - C_\sigma(\hat{A}_i)-C_\tau(\hat{A}_i)]}\,,
\label{cobs}
\end{eqnarray}
where $C_\sigma(\hat{A}_i)$ ($C_\tau(\hat{A}_i)$) denote the total
number of $\sigma$-clusters ($\tau$-clusters) on the Riemann surface
with cut $\hat{A}_i$. 
Since the clusters are non local objects, they represent  ``improved'' observables and the variance 
for the Monte Carlo history of Eq. (\ref{cobs}) is much smaller than in the naive implementation. 

\section{The entanglement entropy in the Ashkin-Teller model}
\label{ATres}

\begin{figure}[t]
\begin{center}
\includegraphics[width=.8\textwidth]{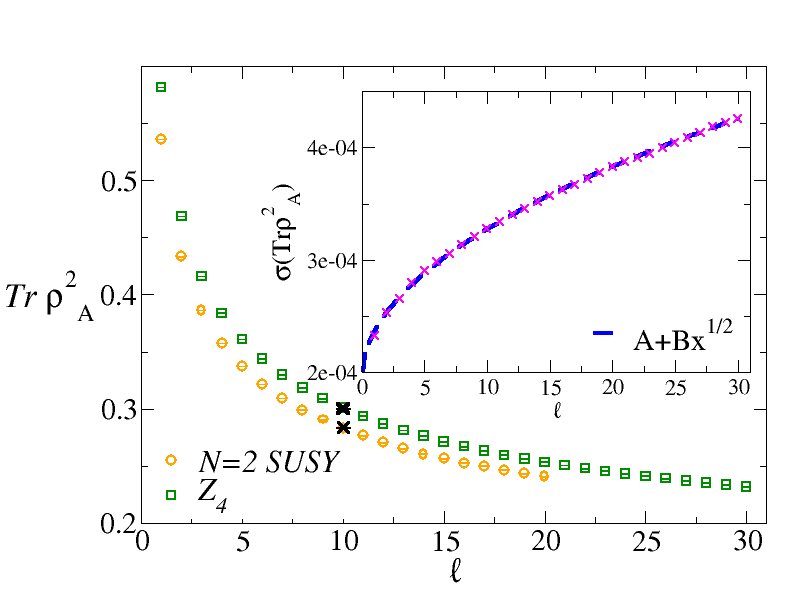}
\end{center}
\caption{$\Tr\rho_A^2$ for a single interval of length $\ell$ in a finite system of length $L=120$. 
Data have been obtained by Monte Carlo simulations using the embedded algorithm. 
  The orange points correspond to the SUSY model and the green ones to the $Z_4$ parafermions. 
  The black crosses at $\ell=10$ are data obtained using the direct algorithm. 
  Inset: behavior of the statistical error of $\Tr\rho_A^2$  vs $\ell$ for the SUSY model. 
  The blue-dashed line is the expected form $A+B\ell^{1/2}$.} 
\label{example}
\end{figure}

\subsection{The single interval}

We first present the results for the Ashkin-Teller model for a single interval.
Although these results do not provide any new information about the model, they are fundamental checks 
for the effectiveness of the Monte Carlo algorithms.
We performed simulations using both algorithms described in the previous section: the direct cluster algorithm and the embedded one. 
When using the direct algorithm, measures are performed using the observable (\ref{cobs}), 
while for the embedded algorithm we used the observable in Eq.  (\ref{obs}).
In Fig. \ref{example} we report the results of the simulations of $\Tr\rho_A^2$ 
for the SUSY model ($r_{\rm orb}=\sqrt{3}/2$ in Fig. \ref{fig14})
and for the $Z_4$ parafermions ($r_{\rm orb}=\sqrt{3/2}$) both for $L=120$. 
The orange and green points are obtained using the embedded algorithm. 
To check  the implementation of the cluster observable, we report at $\ell=10$ the data obtained
using the direct algorithm and Eq. (\ref{cobs}). 
The perfect agreement between the two results confirms the correctness of both implementations.
Note that  $\Tr\rho_A^2$ is a monotonous function of $\ell$, in contrast with the parity effects found for the XXZ spin chain
\cite{ccen-10,ce-10} that also corresponds to a vertex model \cite{BB}.   
In the inset we show the behavior of the statistical error of the observable (\ref{obs}) in the  SUSY case
as function of the subsystem length $\ell$. 
It agrees with the  prediction in Eq.  (\ref{error}) and its absolute value is extremely small, smaller than the size of the points 
in the main plot in Fig. \ref{example}.
Analogous results have been obtained for all the critical points on the self-dual line using both algorithms.

\begin{figure}[t]
\begin{center}
\includegraphics[width=.8\textwidth]{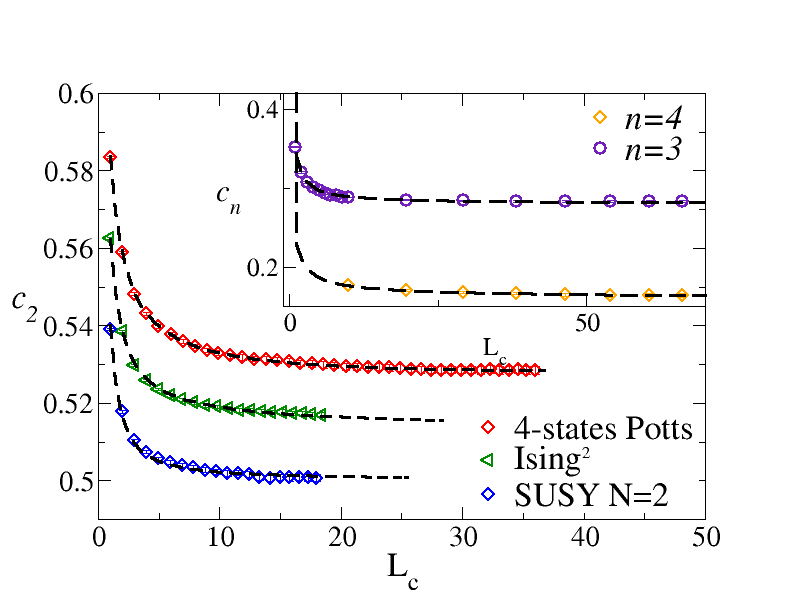}
\caption{Plot of $c_2(L_c)$  as function of $L_c$ for different $\ell$ and $L$.
Three points on the self-dual line are reported: four-states Potts model, uncoupled Ising, and SUSY. 
The dashed lines are fits   to the function $c_2+BL_c^{-K_L}$ ($c_2+AL_c^{-K_L}+BL_c^{-2K_L}$ for the 4-states Potts model) where
 $K_L$ is $1/2,1,4/3$ respectively for the four-states Potts model, Ising, and SUSY. 
  In the   inset we report $c_n$ for $n=3,4$ for the SUSY point. 
  The   dashed lines are fit to $A+BL_c^{2K_L/n}$, with $K_L=4/3$ fixed.}   
\label{c2}
\end{center}
\end{figure}

The results for $\Tr\rho_A^2$ in a finite system are asymptotically described by the CFT prediction (\ref{SnFS}) with $n=2$ and $c=1$.
It is then natural to compute the ratio 
\be
c_2(L_c)=  \frac{\Tr\rho_A^2}{(\frac{L}{\pi}\sin(\frac{\pi}{L}\ell))^{-1/4}}\,,
\label{c2L}
\ee
that is expected to be asymptotically a function of the chord-length 
$L_c=[\frac{L}{\pi}\sin(\frac{\pi}{L}\ell)]$.
This allows to extract the non-universal quantity $c_2$ and to check the form of the corrections to the scaling. 
In Fig. \ref{c2} we report the results for $c_2(L_c)$ 
for the  SUSY point, for the two uncoupled Ising models, and for the four states Potts model. 
It is evident that  for large $L_c$, $c_2(L_c)$ approaches a constant value around $0.5$. 
This is a first confirmation of the CFT predictions on the self-dual line. 

The previous results also provide a test for the theory of the corrections to the scaling to $S^{(n)}_A$.
It has been shown \cite{ccen-10,ce-10} that for gapless models described by a Luttinger liquid theory, the corrections to
the scaling have the form $\ell^{-2K_L/n}$ (or $L_c^{-2K_L/n}$ for finite systems)
where $K_L$ is the Luttinger parameter, related to the circle compactification radius  $K_L=1/2\eta$. 
On the basis of general CFT arguments \cite{cc-10}, it has been argued that this scenario is valid for any CFT
and so also for the AT model with $K_L$ replaced by the dimension of a proper operator.
It is then natural to expect that for the AT model this dimension is $K_L$ in Eq. (\ref{etaAT}), also on the basis of 
the results for the Ising model \cite{ccen-10,ij-08}. 
The dashed lines in Fig. \ref{c2} are fits of $c_2(L_c)$ with the function $c_2+A L_c^{-K_L}$.
The agreement is always very good, except for the four-state Potts model, for which
the exponent of the leading correction $K_L$ assumes the smallest value and so
subleading corrections enter (as elsewhere in similar circumstances, see e.g. \cite{ce-10}). 
In fact, the fit with the function  $c_2+A L_c^{-K_L}+B L_c^{-2K_L}$ is in perfect agreement with the data (but the presence 
of another fit parameter makes this result not so robust). 
This analysis confirms that $K_L$ is the right exponent governing the corrections to the scaling.

\begin{figure}[t]
\includegraphics[width=\textwidth]{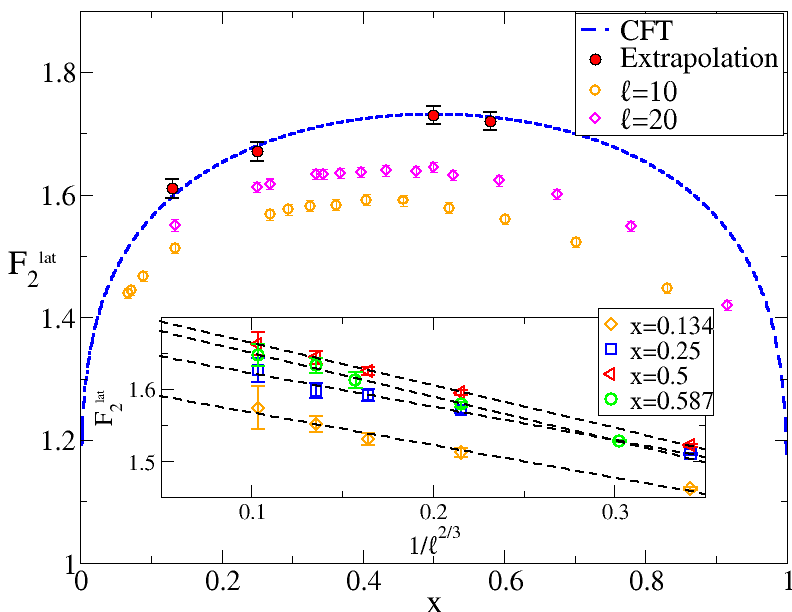}
\caption{
$F_2^{\rm}(x)$ versus the four point ratio $x$ for the SUSY model. 
  The red points are extrapolations obtained using the finite-size ansatz (\ref{ansatz}). 
  The blue-dashed line is the $CFT$ prediction. 
Inset: $F_2^{\rm lat}(x)$ vs $1/\ell^{-2/3}$  for the four values of $x$ used in the extrapolation ($x=0.134,0.25,0.5,0.587$). 
The dashed lines are fits to finite-size ansatz (\ref{ansatz}). }
\label{Susy_F2}
\end{figure}

In the inset of Fig. \ref{c2} we also report  the values of $c_n$ for $n=3,4$ as a function of $L_c$. 
$c_n$ becomes smaller as $n$ increases as for the $XXZ$  \cite{ccen-10},
XX \cite{jk-04}, and Ising \cite{ij-08,ccd-08} spin-chains. 
The dashed lines are fits to the expected scaling behavior $L_c^{-2K_L/n}$ of the corrections, that reproduce perfectly the data.

\subsection{The entanglement entropy of two disjoint intervals.}

In this section we investigate the entanglement entropy of two disjoint intervals and check the correctness of our prediction 
(\ref{atf2}) for the AT model on the self-dual line.  
As for all other cases studied so far numerically (i.e. Heisenberg \cite{fps-08}, Ising \cite{atc-09,fc-10}, and XY \cite{fc-10} chains), 
strong scaling corrections affect the determination of the scaling function $F_n(x)$.
CFT predictions have been confirmed only using the general theory of corrections to the scaling  \cite{ccen-10,ce-10,cc-10,ccp-10}.

In order to determine the function $F_n(x)$, we consider the ratio 
\be
F^{\rm lat}_n(x)= \frac{\Tr \rho_{A_1\cup A_2}^n}{\Tr \rho_{A_1}^n \Tr \rho_{A_2}^n} (1-x)^{c(n-1/n)/6 }\,,
\label{Flat}
\ee
and, on the basis of the general CFT arguments \cite{cc-10}, we expect that the 
the leading correction to scaling can be effectively taken into account by the scaling ansatz
\begin{eqnarray}
F_n^{\rm lat}(x)=F_n^{\rm CFT}(x)+\ell^{-2\omega/n}f_n(x)+\dots\,.
\label{ansatz}
\end{eqnarray}
For the Ising model it has been found $\omega=1/2$ \cite{atc-09,fc-10}. 
Since for $\eta=2$ the AT Hamiltonian reduces to two uncoupled Ising models, one naively expects  
$\omega=K_L/2$ along the whole self-dual critical line of the AT model.

\begin{figure}[t]
\begin{center}
\includegraphics[width=.7\textwidth]{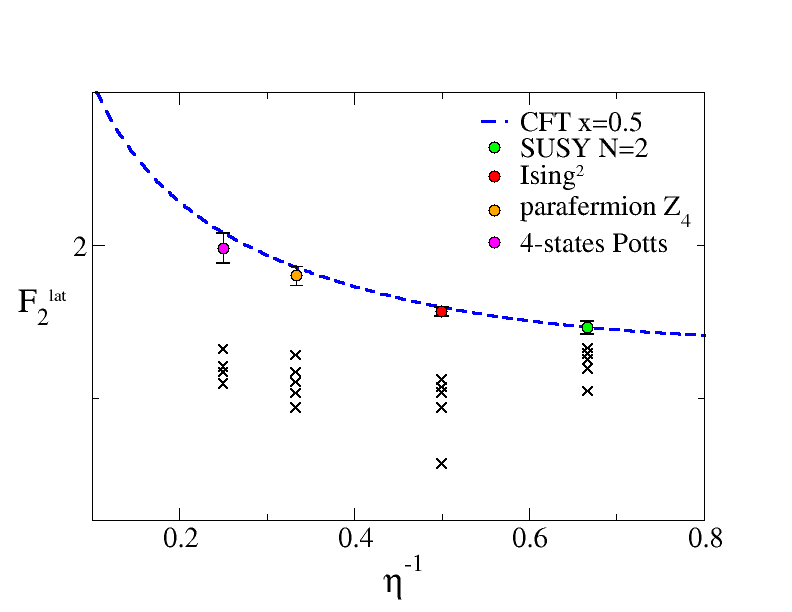}
\end{center}
\caption{$F_2^{\rm lat}(1/2)$ as function of $\eta^{-1}$ for different models (Ising, SUSY, $Z_4$ parafermions, and four-states Potts model). 
  The blue-dashed line is the CFT prediction. 
  The (colored) points close to the curve are extrapolations obtained with the finite-size scaling ansatz (\ref{ansatz}). 
  The  black crosses are the Monte Carlo data used for the fits. 
  The block lengths used range from $\ell=5$ to $\ell=80$.}
\label{all}
\end{figure}


Hereafter we only consider $\Tr\rho^2_{A}$. 
We start our analysis from the SUSY point  that (assuming $\omega=K_L/2$) should have the smaller corrections to scaling.
In Fig. \ref{Susy_F2} we show Monte Carlo data at $\ell=10,20$ ($L=120$) for $F_2^{\rm lat}(x)$ plotted against the 
four point ratio $x$ defined as in Eq. (\ref{xFS}). 
We report with the blue dashed line the asymptotic CFT result  (cf. Eq.  (\ref{atf2})). 
As in all other cases considered in the literature \cite{atc-09,fc-10}, the curves for $F_2(x)$ at $\ell=10,20$ are not 
symmetric functions of $x\to1-x$, as instead the asymptotic CFT prediction must always be \cite{fps-08}. 
This is due to the non-symmetrical finite-size corrections $f_2(x)$ in Eq. (\ref{ansatz}).  
We extrapolate the result at $\ell\to\infty$ using the ansatz (\ref{ansatz}) and  $\omega=2/3$.
The extrapolations are reported as red points in Fig.  \ref{Susy_F2}.
There is a very good agreement between  the extrapolations and the theoretical curve. 
Since the correction exponent $\omega=2/3$ is rather large, and so the corrections small, 
even small subsystems such as $\ell=10,20$ are  enough to obtain a good extrapolation.
In  the inset of Fig. \ref{Susy_F2} we report the Monte Carlo data for $F^{\rm lat}_2(x)$ against $\ell^{-2/3}$.
The linear behavior in this inset confirms the validity of the ansatz (\ref{ansatz}) and the reported 
 straight lines are the fits giving the extrapolations reported in the main panel.

\begin{figure}[t]
\includegraphics[width=.6\textwidth]{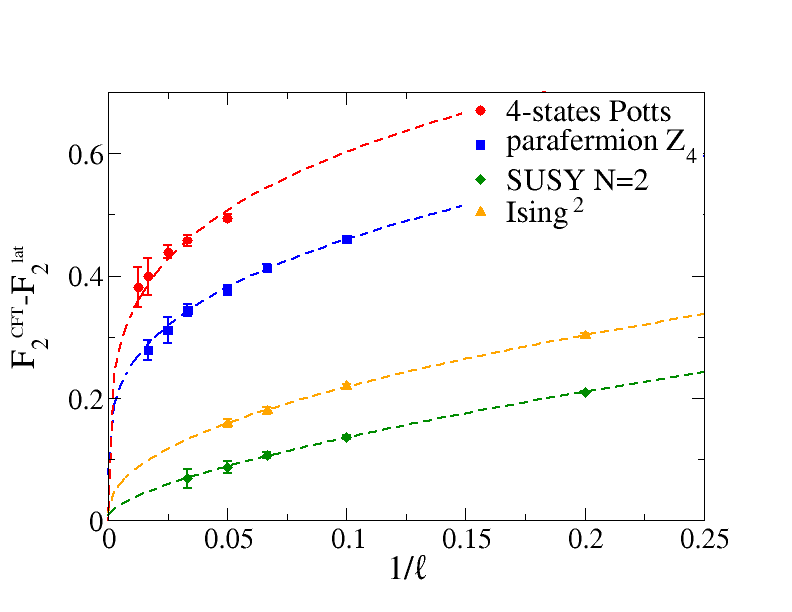}
\includegraphics[width=.6\textwidth]{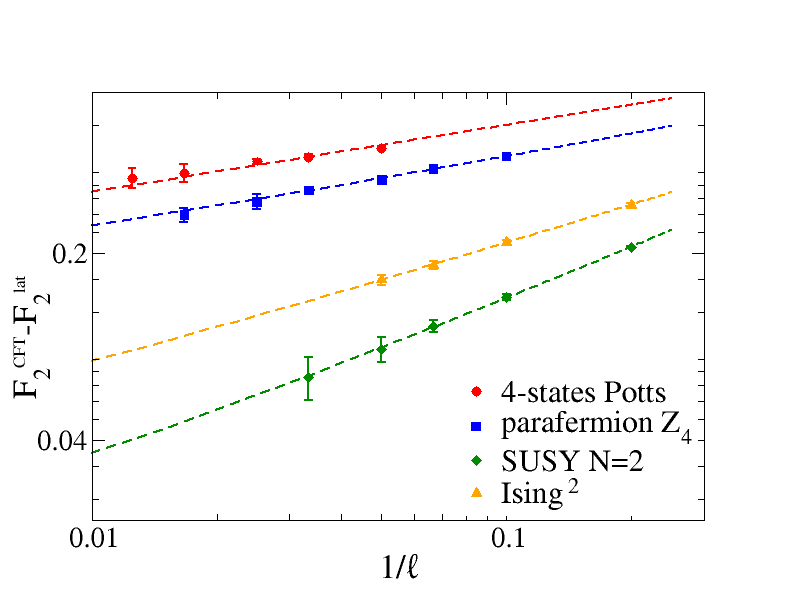}
\caption{$F_2^{\rm CFT}(1/2)-F_2^{\rm lat}(1/2)$ versus $1/\ell$. 
 The dashed lines are fits to the finite-size scaling ansatz (\ref{ansatz}) fixing the value of $F_2^{\rm CFT}(1/2)$.
  Left: the same plot  in log-log scale.   }
\label{fits}
\end{figure}

We also investigate other points on the self dual line, namely the $4$-states  Potts model ($\eta=4$), the
parafermion $Z_4$ ($\eta=3$), the uncoupled Isings ($\eta=2$). 
In Fig.~\ref{all} we report $F^{\rm lat}_2(x)-F_2^{\rm CFT}(x)$ at fixed $x=1/2$ versus $\eta^{-1}$ for all the mentioned models.
we report $x=1/2$ because it is the value of $x$ providing the most stable estimate, but also other values have been studied. 
Indeed, on one hand, the computational cost of the simulations decreases going toward $x=1$ (the reason being evident from
the definition of  $x$ for which smaller lattice sizes are needed). 
On the other hand, scaling corrections become more severe in the region $x\sim 1$, as clear from the results for the SUSY model
in Fig. \ref{Susy_F2}. 
Thus $x=1/2$ represents the best compromise between these two drawbacks. 
The dashed curves in the left panel of Fig.~\ref{fits} are fits of the data with Eq. (\ref{ansatz}) obtained by fixing the value of 
$F_2^{\rm CFT}(x)$ to its predicted value (cf. Eq.  (\ref{atf2})). 
There is a very good agreement with the full theoretical picture, confirming in particular the correctness of the exponent governing 
the leading correction to the scaling. 
For the $Z_4$ parafermions and for the four-state Potts model, we needed very large values of $\ell$ in order to
show the correct asymptotic behavior (the range of $\ell$ reported in the plot is  in fact $5\le\ell\le80$).
This is made clearer in the right panel of Fig. \ref{fits} where the same data are shown in log-log scale.
In  Fig. \ref{all}   we reports the fits obtained by
fixing only the exponent of the corrections $\omega=K_L/2$ and leaving $F_2^{\rm CFT}(1/2)$ free. 
For all considered values of $\eta$, the extrapolation of $F^{\rm lat}_2(1/2)$ to $\ell\to\infty$ is compatible (within error bars) 
with the expected result $F_2^{\rm CFT}(1/2)$.

We finally study the correction amplitude $f_2(x)$ in Eq. (\ref{ansatz}).
This function is the main reason of the asymmetry in $x\to1-x$ for $F_2^{\rm lat}(x)$ and knowing its gross features
could greatly simplify future analyses. 
For the Ising model, it has been found that $f_2(x)\sim x^{1/4}$ for small $x$, that is the same behavior of $F_2(x)-1$. 
Since along the whole self-dual line $F_2(x)-1\sim x^{1/4}$, we would expect 
\be
f_2(x)\sim  x^{1/4}\,.
\label{f2hyp}
\ee
For the Ising model (i.e. $\eta=2$), this scenario has been already verified with high precision \cite{atc-09}.

\begin{figure}[t]
\begin{center}
\includegraphics[width=.7\textwidth]{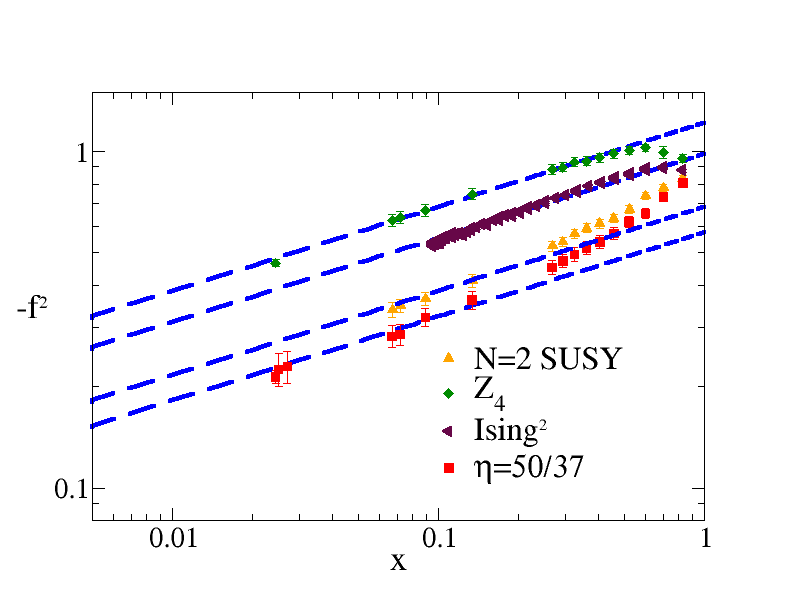}
\end{center}
\caption{Monte Carlo data for $f_2(x)$ obtained as
  $f_2(x)=(F^{\rm lat}_2(x)-F_2^{CFT}(x))\ell^{K_L/2}$ as function of $x$. We show data for
  $\ell=10$ and various models (SUSY, $Z_4$ parafermions,
  Ising model and the model corresponding to $\eta^{-1}=0.74$). The
  blue-dashed lines are asymptotic fits to $Ax^{1/4}$.}
\label{sub}
\end{figure}

In Fig. \ref{sub}  we report $f_2(x)$ obtained as $f_2(x)=(F_2^{\rm CFT}(x)-F_2^{\rm lat})\ell^{K_L/2}$
as function of  $x$ (in logarithmic scale to highlight the small $x$ behavior). 
All data correspond to $\ell=10$ and various values of $L$.
For the two  largest values of $\eta$ ($Z_4$ parafermionic theory at $\eta=3$ and for the Ising  model at $\eta=2$),
we observe an excellent agreement with our conjecture $f_2(x)\sim x^{1/4}$.
However decreasing the value of $\eta$, i.e. for the SUSY model at $\eta=3/2$ and for the model at $\eta^{-1}=0.74$, 
the behavior of $f_2(x)$ is not as linear as before, especially for high value of $x$.
Nonetheless for $x<0.4$ the data  confirm the behavior $x^{1/4}$. 
Furthermore, it seems that for any $\eta\neq2$, subleading terms in the expansion for small $x$ appear
and they are vanishing only for the Ising model. 

\section{The Tree Tensor Network}
\label{ttn:sec}

This section is divided into two parts. 
First we explain in a self contained way how to extract the spectrum of the reduced density matrix of 
some specific bipartitions of a pure state encoded in a Tree Tensor Network (TTN).
We only recall the basic definitions introduced in Ref. \cite{TTN} and refer the reader to the literature for complementary 
works on the subject \cite{fnw-92,f-97,hieida,lcp-00,mrs-02,sdv-06,nagaj-08,silvi,Dur,Murg,Plenio,Gliozzi,gauge}.
Secondly we quickly recall how to use TTN to calculate the ground state of the anisotropic Heisenberg spin-chain.

\subsection{Tree Tensor network and reduced density matrices.}

\begin{figure}[t]
\begin{center}
\includegraphics[width=0.8\textwidth]{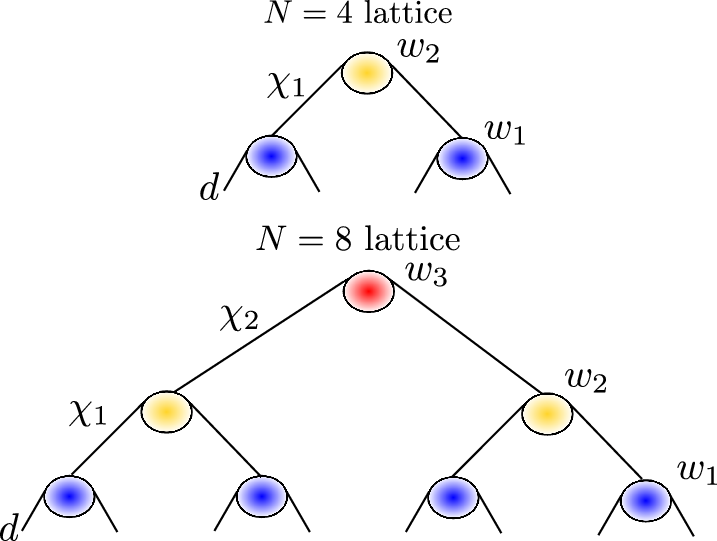}
\caption{Examples of TTN for a $N=4$ lattice and a $N=8$ lattice.} 
\label{fig:SmallTree}
\end{center}
\end{figure}

We consider a one dimensional lattice $\mathcal{L}$ made of $N $ sites, where each site is described by a local Hilbert 
space $\mathbf{V}$ of finite dimension $d$. 
In this work the state is  the  ground state $|\Psi_{\mbox{\tiny GS}}\rangle$ of some local Hamiltonian $H$ defined on $\mathcal{L}$, 
but in general it could be an arbitrary pure state $|{\Psi}\rangle \in \mathbf{V}^{\otimes N}$ defined on the lattice $\mathcal{L}$. 

A generic state $|\Psi\rangle\in \mathbf{V}^{\otimes N}$ can always be expanded as
\begin{eqnarray}
|\Psi\rangle = \!\sum_{i_1=1}^d ~ \sum_{i_2=1}^d \cdots \sum_{i_N=1}^d T_{i_1i_2 \cdots i_N} 
|  i_1\rangle|  i_2 \rangle \cdots | i_N \rangle,
\label{eq:local_expansion}
\end{eqnarray}
where the $d^{N}$ coefficients $T_{i_1i_2 \cdots i_N}$ are complex numbers and the vectors 
$\{| 1_s \rangle, |2_s\rangle, \cdots, |d_s\rangle \}$ denote a local basis on the site $s\in \mathcal{L}$. 
We refer to the index $i_s$ that labels a local basis for site $s$ ($i_s=1,\cdots,d$) as a \emph{physical} index. 

In the case we are interested in, the tensor of coefficients $T_{i_1i_2 \cdots i_N}$ in Eq. (\ref{eq:local_expansion}) is the result 
of the contraction of a TTN. 
As shown in Fig. \ref{fig:SmallTree} for lattices of $N=4$ and $N=8$ sites, a TTN decomposition of  $T_{i_1i_2 \cdots i_N}$ consists 
of a collection of tensors $w$ that have both \emph{bond} indices and \emph{physical} indices. 
The tensors are interconnected by the bond indices according to a tree pattern. 
The $N$ physical indices correspond to the leaves of the tree. 
Upon summing over all the bond indices, the TTN produces the $d^N$ complex coefficients $T_{i_1i_2 \cdots i_N}$ of 
Eq. (\ref{eq:local_expansion}).

\begin{figure}[t]
\begin{center}
\includegraphics[width=8cm]{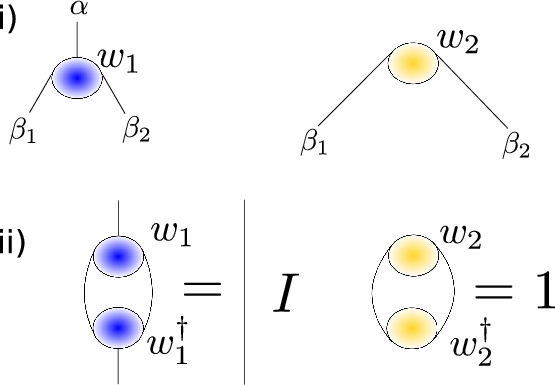}
\caption{ (i) Diagrammatic representation of the two types of isometric tensors in the TTN for a $N=4$ lattice in Fig. \ref{fig:SmallTree}. 
(ii) Graphical representation of the constraints in Eqs. (\ref{eq:const1}) and (\ref{eq:const3}) fulfilled by the isometric tensors. } 
\label{fig:Isometric}
\end{center}
\end{figure}

The tensors in the TTN will be constrained to be \emph{isometric}, in the following sense. 
As shown in Fig. \ref{fig:Isometric} for the $N=4$  lattice of Fig. \ref{fig:SmallTree}, each tensor $w$ in a TTN has at 
most one upper leg/index $\alpha$ and two  lower indices/legs $\beta_1, \beta_2$, so that its entries read 
$(w)^{\alpha}_{\beta_1,\beta_2}$ (everything can be generalized to tensors with more upper and lower legs \cite{TTN}). 
Then we impose that
\be
\sum_{\beta_1 , \beta_2} (w)_{\beta_1 , \beta_2}^{\alpha}(w^{\dagger})^{\beta_1 , \beta_2}_{\alpha'} = \delta_{\alpha\alpha'}.
	\label{eq:isometry}
\ee
For clarity, throughout this paper we use diagrams to represent tensors networks as well as tensor manipulations. 
For instance, the constraints for the tensors $w_1$ and $w_2$  of the TTN of Fig. \ref{fig:SmallTree} for a $N=4$ lattice, namely
\begin{eqnarray}
\sum_{\beta_1 \beta_2 } (w_1)_{\beta_1 \beta_2}^{\alpha} (w_1^{\dagger})^{\beta_1 \beta_2}_{\alpha'} &=& \delta_{\alpha\alpha'}, \label{eq:const1}\\
	\sum_{\beta_1 \beta_2} (w_2)_{\beta_1 \beta_2}(w_2^{\dagger})^{\beta_1 \beta_2} &=& 1,\label{eq:const3}
\end{eqnarray}
are represented as the diagrams in Fig. \ref{fig:Isometric}(ii). 
We refer to a tensor $w$ that fulfills Eq. (\ref{eq:isometry}) as an \emph{isometry}.

\begin{figure}[t]
\begin{center}
\includegraphics[width=8cm]{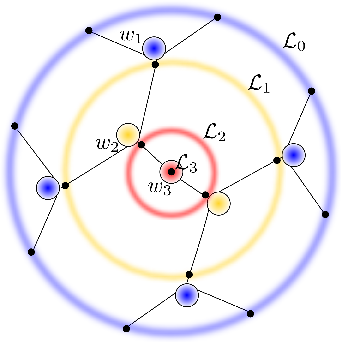}
\caption{The isometric TTN of Fig. \ref{fig:SmallTree} for a $N=8$ lattice $\mathcal{L}_0$ with periodic boundary conditions 
(the blue external circle) is associated with a coarse-graining transformation that generates a sequence of increasingly 
coarse-grained lattices $\mathcal{L}_1$, $\mathcal{L}_2$ and $\mathcal{L}_3$ (the inner circles). 
Notice that in this example we have added an extra index to the top isometry $w_{3}$, corresponding to the single site of an 
extra top lattice $\mathcal{L}_3$, which we can use to encode in the TTN a whole subspace of $\mathbf{V}^{\otimes N}$ 
instead of a single state $|\Psi\rangle$.} 
\label{fig:CoarseGrain}
\end{center}
\end{figure}

An intuitive interpretation of the use of a TTN to represent a state $|\Psi\rangle$ can be obtained in terms of a 
coarse-graining transformation for the lattice $\mathcal{L}$. 
Notice that the isometries $w$ in Fig. \ref{fig:SmallTree} are organized in layers. 
The bond indices between two layers can be interpreted as defining the sites of an effective lattice. 
In other words, the TTN defines a sequence of increasingly coarser lattices 
$\{\mathcal{L}_0, \mathcal{L}_1, \cdots, \mathcal{L}_{T-1} \}$, where 
$\mathcal{L}_0 \equiv \mathcal{L}$ and each site of lattice $\mathcal{L}_{\tau}$ is defined in terms of several sites of 
$\mathcal{L}_{\tau-1}$ by means of an isometry $w_{\tau}$, see Fig. \ref{fig:CoarseGrain}. 
In this picture, a site of the lattice $\mathcal{L}_{\tau}$ effectively corresponds to some number $n_{\tau}$ of sites of the original 
lattice $\mathcal{L}_0$. 
For instance, each of the two sites of $\mathcal{L}_{2}$ in Fig. \ref{fig:CoarseGrain} corresponds to $8$ sites of $\mathcal{L}_0$.
Similarly, each site of lattice $\mathcal{L}_{1}$ corresponds to $4$ sites of $\mathcal{L}_0$.

\begin{figure}[t]
\begin{center}
\includegraphics[width=6cm]{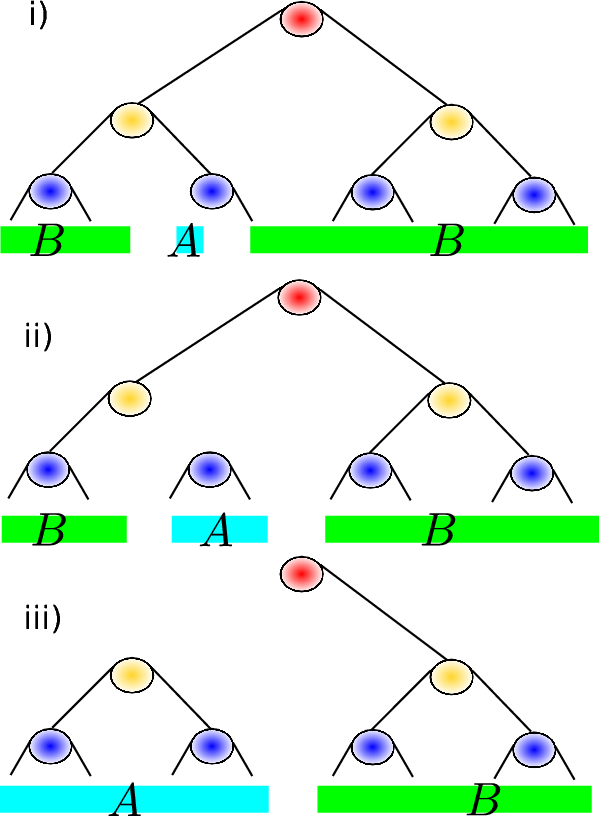}
\caption{By erasing one of the  indices in the TTN the spin chain is always divided in two parts $A$ and $B$ \cite{sdv-06}. 
Here we show that in the case of the $N=8$ lattice of Fig. \ref{fig:SmallTree} there are three classes of indices, identified by 
their position in the TTN. 
i) physical bonds connect a single spin with the rest of the lattice, 
ii) bond indices of the first layer connect a block of  two adjacent spins to the rest of the lattice, 
iii) bond indices of the third layer of the lattice connect four adjacent spins, to the other half. 
This implies that the rank of the index is the Schmidt rank of the respective partition. 
} 
\label{fig:bond}
\end{center}
\end{figure}

The use of isometric tensors, and the fact that each bond unambiguously defines two parts $(A:B)$ of the chain which are connected only through that bond as displayed in Fig. \ref{fig:bond},   implies that the rank of that bond in the TTN is given by the Schmidt rank $\chi(A:B)$ of the  partition $(A:B)$ \cite{sdv-06}. 
Thus the reduced density matrix $\rho_A$ for a set $A$ of sites of $\mathcal{L}$ is
\begin{equation}
\rho_A = \tr_{B} \proj{\Psi} = \sum_{\alpha} p_{\alpha} \proj{\Psi^{A}_{\alpha}},
\label{eq:rhoA}
\end{equation}
where 
$p_{\alpha}$ are the eigenvalues of $\rho_A$.  
It follows then the R\'enyi entanglement entropies $S_A^{(n)}$ are 
\begin{equation}
S_A^{(n)}=\frac1{1-n}\log{\rm Tr}\,\rho_A^n=
\frac1{1-n} \log \sum_{\alpha} p_{\alpha}^n\,,
\ee
and for $n=1$
\be
S_A^{(1)}= -\tr(\rho_A \log \rho_A) = -\sum_{\alpha} p_{\alpha} \log p_{\alpha}.
\label{eq:entropy}
\end{equation}

In the following  we denote the ranks of the  tensor $w_{\tau}$, $\alpha, \beta_1, \beta_2$  
as  $, \chi^{\tau},\chi^{\tau -1}, \chi^{\tau -1}$. In general, they fulfill
\begin{equation}
	 \chi^{\tau} < (\chi^{\tau -1})^2, 
\end{equation}
meaning that  $w_{\tau}$  projects states in $\mathbf{V}^{\tau-1}\otimes \mathbf{V}^{\tau-1}$ into the smaller Hilbert 
space  $\mathbf{V}^{\tau}$.

For a critical chain, the logarithmic scaling of the entanglement entropy (cf. Eq. (\ref{Renyi:asymp}))
implies that  the rank of the isometries should at least grow proportionally to the length of the block represented by the effective spins
\begin{equation}
	 \chi^{\tau} \propto n_{\tau}, 
\end{equation}
which means that while moving to higher layer of the tensor network the rank of the isometries increases. 
This also implies that the leading cost of the computation is concentrated in contracting the first few layers of the TTN.  
If $N=2^T$ and we describe a pure state (so that the rank of the $\alpha_{\tau}$ is one) the maximal rank of the tensors in the  
TTN is 
\begin{equation}
	 \chi=\max_{\tau} \chi ^{\tau}=\chi^{T-1}. 
\end{equation}

\begin{figure}[t]
\begin{center}
\includegraphics[width=9cm]{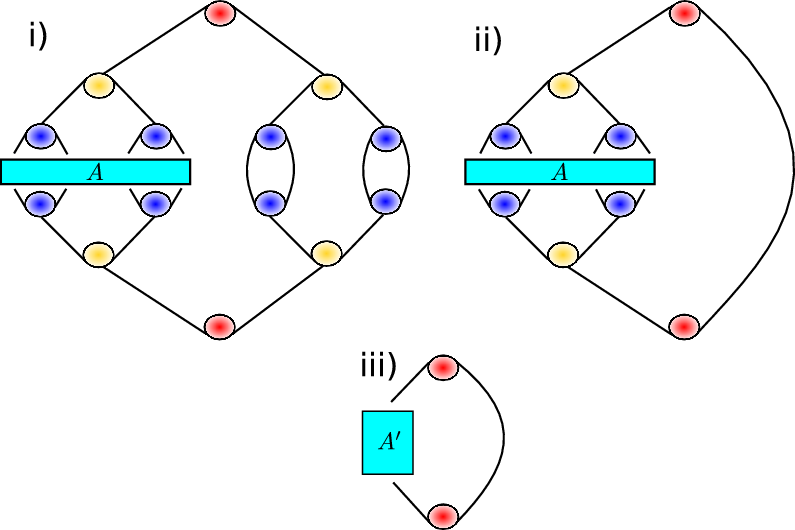}
\caption{Computation of the spectrum $\{p_{\alpha}\}$ of the reduced density matrix $\rho_A$ for a block $A$ that corresponds 
to one of the coarse-grained sites. 
(i) Tensor network corresponding to $\rho_A$ where $A$ is half of the lattice. 
(ii) Tensor network left after several isometries are annihilated with their Hermitian conjugate. 
(iii) since the spectrum of $\rho_A$ is not changed by the isometries acting on $A$, we can  eliminate them and 
we are left with a network consisting of only two tensors, which can now be contracted together. 
The  cost of this computation is 
proportional to $\mathcal{O}(\chi'^3\chi^2)\le \mathcal{O}( \chi^5 )$.} 
\label{fig:SpectEval}
\end{center}
\end{figure}

In Ref. \cite{TTN} it has been shown that 
i) a TTN description of the ground state of chain of length $N$ with periodic boundary conditions can be obtained numerically with 
a cost of order $\mathcal{O}(\log N \chi^4)$.  
ii) From the TTN it is also straightforward to compute the spectrum $\{p_{\alpha}\}$ of the reduced density matrix $\rho_A$ 
(cf. Eq. (\ref{eq:rhoA}))  when $A$ is a block of contiguous sites corresponding to an effective site of 
any of the coarse-grained lattices $\mathcal{L}_1, \cdots, \mathcal{L}_{T-1}$. 
Fig. \ref{fig:SpectEval} illustrates the tensor network corresponding to $\rho_A$ for the case when $A$ is one half of the chain. 
Many pairs of isometries are annihilated. In addition, the isometries contained within region $A$ can be removed since they 
do not affect the spectrum of $\rho_A$. 
From the spectrum $\{p_{\alpha}\}$, we can now obtain the R\'enyi entropies $S_A^{(n)}$.
The leading cost for  computing the spectrum of the reduced density matrix $\rho_A$ for this class of bipartitions is due to
the contractions of the first layers of the TTN. 
When the bipartition is such that $A$ is a quarter of the chain, this  implies a cost  proportional to 
$\mathcal{O}(\chi'^3\chi^2)\le \chi' \chi^4 $, where $\chi'=\chi^{T-2}$.

\begin{figure}[t]
\begin{center}
\includegraphics[width=9cm]{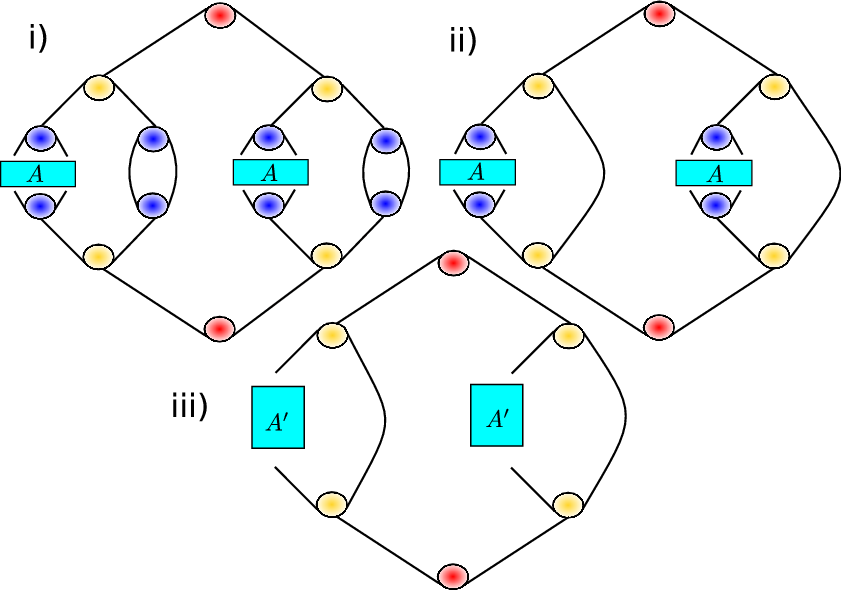}
\caption{Computation of the spectrum $\{p_{\alpha}\}$ of the reduced density matrix $\rho_A$ when $A$  corresponds to two  
coarse-grained sites separated by one coarse grained site from both sides. 
(i) Tensor network corresponding to $\rho_A$ where $A$ is a quarter of the lattice. 
(ii) Tensor network left after several isometries are annihilated with their Hermitian conjugate. 
(iii) Since the spectrum of $\rho_A$ is not changed by the isometries acting on $A$, we can eliminate them and we 
are left with a network consisting of only few tensors, which can now be contracted together.
The cost of contracting this tensor network is proportional to 
$\max[\mathcal{O}(\chi^2 \chi'^4),\mathcal{O}(\chi^3 \chi'^2)] < \mathcal{O}(\chi^6)$.} 
\label{fig:SpectEvaltwoBlocks}
\end{center}
\end{figure}

It is also possible to compute the reduced density matrix $\rho_A$  when $A$ is composed of two disjoint  subintervals 
$A_1$ and $A_2$, 
where now each of the two intervals is a block of contiguous sites corresponding to an effective site of the coarse grained lattice. 
The cost  of this computation is again dominated by contracting the upper part of the tensor network, and the most expensive case is
obtained by considering $A$ as  the collection of two $N/4$ spins blocks, separated by $N/4$ spins. 
The tensor network corresponding to this $\rho_A$ is shown in Fig. \ref{fig:SpectEvaltwoBlocks}. 
Also in this case many pairs of isometries are annihilated. The isometries contained within the composed region $A$ can also 
be removed since they do not affect the spectrum of $\rho_A$.  
The cost of contracting this tensor network is proportional to 
$\max [\mathcal{O}(\chi^2 \chi'^4), \mathcal{O}(\chi^3 \chi'^2)] < \mathcal{O}(\chi^6)$. 

\subsection{The TTN and the anisotropic Heisenberg spin-chain}

In the previous subsection we have shown how to extract the spectrum of the reduced density matrix for a single and a double 
spin block from a TTN state.
In this manuscript we are interested in reduced density matrices calculated on the ground-state of the anisotropic 
Heisenberg spin chain (XXZ model) in zero magnetic field, defined by the Hamiltonian
\be
H=\sum_{j=1}^L [\s^x_j\s^x_{j+1}+\s^y_j \s^y_{j+1}+\Delta
\s^z_j\s^z_{j+1}]\, , 
\label{HXXZ}
\ee
where $\s_j^\alpha$ are the Pauli matrices at the site $j$. Periodic boundary conditions are assumed. 
We are interested in gapless conformal phases of the model, that is $-1<\Delta\leq 1$. 
This phase is described  by a free-bosonic CFT compactified on a circle with radius that depends on the parameter $\Delta$ 
\be
\eta=2r_{\rm circle}^2=\frac1{2K_L}=\frac{\arccos (-\Delta)}\pi\,,
\label{etaDe}
\ee
where $K_L$ is the Luttinger liquid parameter. \footnote{Notice similarities and differences between Eq. (\ref{etaDe}) and  
its analogous for the AT model (\ref{etaAT}). The relation between $\eta$ and $r^2$ and the relation between $K_L$ and 
$\Delta$ are the same for both XXZ spin-chain and AT model, but the relation between $\eta$ and $\Delta$ (or $K_L$ and $r$)
is different.
}
The sign convention in the Hamiltonian (\ref{HXXZ}) is such that the model is (anti)ferromagnetic for $\Delta<0$ ($\Delta>0$).
Hamiltonian (\ref{HXXZ}) is diagonalizable by means of Bethe ansatz.
However, obtaining the spectrum of the reduced density matrix from Bethe ansatz is still a major problem 
and only results for small subsystems are known \cite{afc-09,ncc-09}.
For this reason we exploit variational TTN techniques to obtain the ground state. 

Here we follow the variational procedure described in detail in Ref. \cite{TTN}, 
where the generic technique (consisting of assuming a tensor network
description of the ground state and minimize the energy variationally
improving the tensors one by one as described, i.e., in Ref. \cite{cv-09}) has been specialized and optimized for the case of a TTN.  
We exploit translation invariance by using the same tensor at each layer of the TTN. 
One could also improve the efficiency further by exploiting the $U(1)$ symmetry of the Hamiltonian (\ref{HXXZ}), i.e. the
rotations around the $z$ axis. However we did not make use of this symmetry here.

\section{The Block Entanglement of the Anisotropic Heisenberg spin-chain}
\label{XXZ:sec}

\begin{figure}[t]
\begin{center}
\includegraphics[width=.8\textwidth]{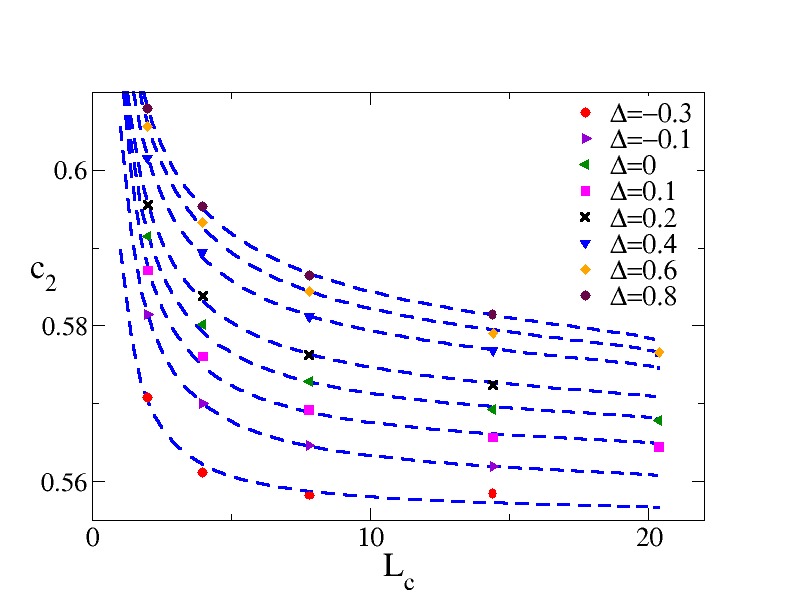}
\end{center}
\caption{TTN data for the non universal constant $c_2(L_c)$ as function of the chord length $L_c$ 
for different values of $\Delta$. The dashed curves are fits
  to the function $A+BL_c^{-K_L}$. The reported data have been
  obtained with $L=128$ for $\Delta=0,0.1,0.6$ and $L=64$ for the 
  other values.}
\label{XXZsingle}
\end{figure}

In this section we report the TTN results for the R\`enyi entropies in the XXZ spin-chain for a single and a double interval. 
As a main advantage compared to the classical Monte Carlo simulations performed for the AT model, with a single TTN simulation we
obtain the spectrum of the reduced density matrix and hence any R\`enyi entropy, including von Neumann $S_A^{(1)}$.
Oppositely with the Monte Carlo methods only R\`enyi entropies $S_A^{(n)}$ of integer order $n\geq2$ 
can be obtained and each of them  requires an independent simulation.

\subsection{The single interval.}

We first present the TTN results for the single interval. 
These have been already obtained with many numerical variational techniques \cite{ccen-10,lsca-06,osc,xa-11}
and are reported here only to test the accuracy of the TTN and to fix units/scales etc. 
Using variational TTN, we find the ground-state of the XXZ Hamiltonian (\ref{HXXZ}) and from this we extract the 
spectrum of the reduced density matrix of the single block, as explained in the previous section.  
We then numerically obtain $\Tr\rho_A^n$.
The maximum size of the chain that we consider is $L=128$.
The subsystem lengths considered are $\ell=2,4,8,16,32$. 
Notice that with the TTN method, using a binary tree as we are doing, we can effectively access only subsystems sizes of the form 
$2^m$ with $m$ arbitrary integer, as it should be clear from the previous section.
In particular this limits the calculation to even values of $\ell$ and we can not study the parity effects reported 
in Ref. \cite{ccen-10,ce-10}.

We considered different values of the anisotropy parameter $\Delta$, namely $\Delta=-0.3,-0.1,0,0.1,0.2,0.4,0.6,0.8,1$. 
The TTN becomes less effective for values of $\Delta\leq-0.5$. This can be easily traced back to the smallness 
of the finite-size gap that in the minimization process causes the algorithm to be stuck in meta-stable states when the 
system size is large enough. This drawback could be cured by using larger values of $\chi$ (and so larger computational cost), but 
as we shall see, the considered values of $\Delta$ suffice to draw a very general picture of the entanglement. 
For the isotropic Heisenberg antiferromagnet at $\Delta=1$ we ignore the presence of logarithmic corrections
to the scaling \cite{lsca-06,cc-10}, that have a minimal effect for all our aims.

\begin{figure}
\begin{center}
\includegraphics[width=.8\textwidth]{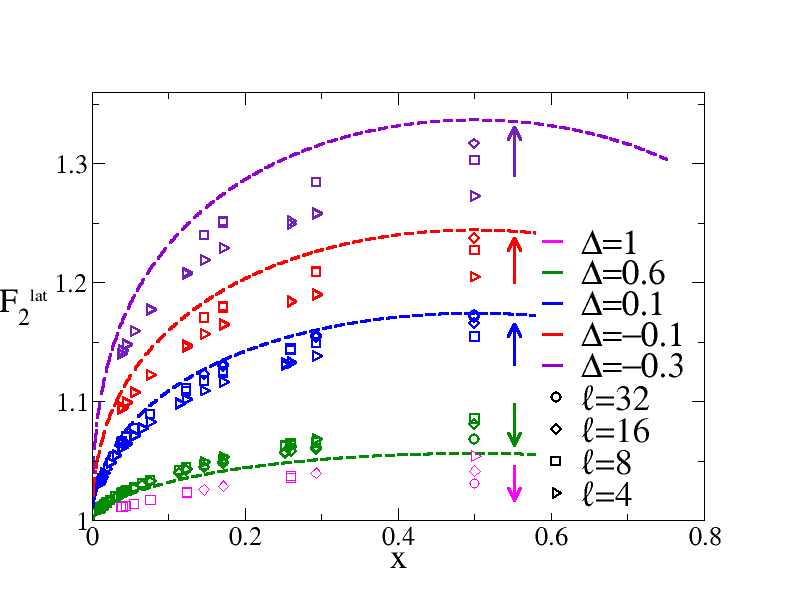}
\end{center}
\caption{TTN data for $F^{\rm lat}_2(x)$ as function of $x$ for various sizes of the chain  $L=16,32,64,128$,  subsystem lengths
  $\ell=4,8,16,32$, and  $\Delta=-0.3,-0.1,01,0.6$. 
  Different values of $\Delta$ are distinguished by different colors, while different symbols denote different
  values of $\ell$. The arrows denote the (asymptotically) increasing subsystem sizes $\ell$. }   
\label{XXZtr_2}
\end{figure}

As for the AT model, we study the quantity $c_2(L_c)$ defined by the ratio in Eq. (\ref{c2L}).
The results are shown in Fig. \ref{XXZsingle} for all considered values of $\Delta$.
The scaling corrections are evident, especially for larger values of $\Delta$, as expected \cite{ccen-10}. 
These corrections for ${\rm Tr}\rho_A^n$ are indeed of the form $L_c^{-2K_L/n}$ \cite{ccen-10} ($K_L$ is defined in Eq. (\ref{etaDe})).
The dashed lines reported in Fig.~\ref{XXZsingle} are fits to this form for $n=2$, showing the 
agreement between  TTN data and the fits.
We checked that all the TTN data agree with the ones obtained in Ref. \cite{ccen-10} using density matrix renormalization group.
The agreement is perfect and for this reason we refer to the above paper for a detailed study of $\Tr\rho_A^n$ for $n>2$.


\subsection{Double interval: the $n=2$ case.}

We now  consider a subsystem made of two parts $A_1$ and $A_2$ of equal length $\ell$. 
We start by studying the quantity $\Tr\rho^2_{A_1\cup A_2}$ for finite chains and extract the
universal function $F_2^{\rm CFT}(x)$ by proper extrapolation. 
Since we only consider even $\ell$, corrections to the scaling are expected to be monotonic in $\ell$ also for $F_2(x)$, 
oppositely to the case of arbitrary $\ell$ parity \cite{fps-08,fc-10}.
The CFT prediction for the function $F_2(x)$ for the XXZ chain is Eq. (\ref{F2}) with $\eta$ given by Eq. (\ref{etaDe}).

\begin{figure}[t]
\begin{center}
\includegraphics[width=.8\textwidth]{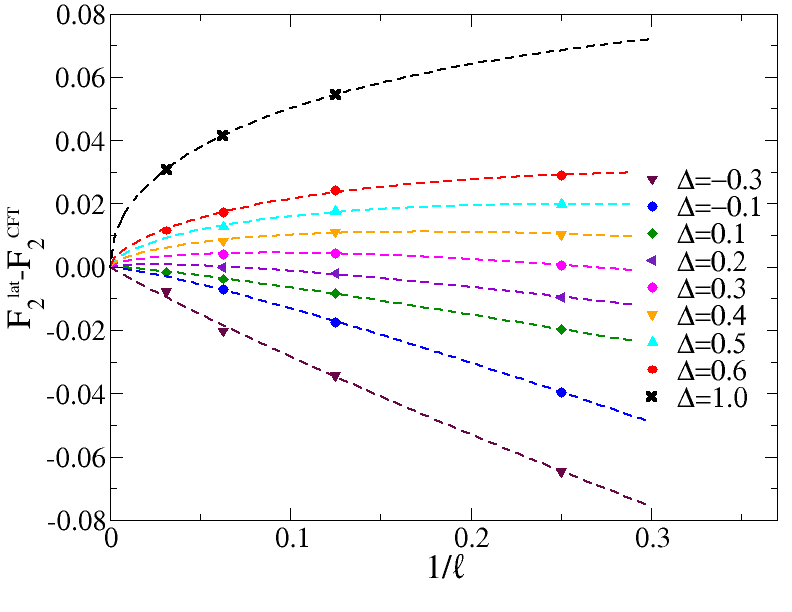}
\end{center}
\caption{TTN data for $F_2^{\rm lat}(1/2)-F_2^{\rm CFT}(1/2)$ as function of $1/\ell$ for various
  $\Delta$. The dashed lines are fits to the function with the generalized finite-$\ell$ ansatz (\ref{ansatz2}).
  }   
\label{XXZfits}
\end{figure}

In Fig. \ref{XXZtr_2} we report TTN data for $F^{\rm lat}_2(x)$ (obtained with the ratio defined in Eq. (\ref{Flat})) as function of the cross
ratio $x$ for  $\Delta=-0.3,-0.1,0.1,0.6$ and subsystem sizes
$\ell=4,8,16,32$.  
The different values of $\Delta$ are denoted with different colors, while the different symbols stand for the various $\ell$. 
On the same figure we also show the asymptotic $F_2^{\rm CFT}(x)$ as dashed lines. 
It is evident that strong scaling corrections affect the data, as expected. 
Colored arrows denote the direction of (asymptotically) increasing subsystem sizes. 
Very surprisingly, while for $\Delta=-0.3,-0.1,0.1$ the asymptotic CFT result is approached from below, 
for $\Delta=0.6$ it is approached from above.  
Moreover, for $\Delta=0.6$ the behavior of the data is not monotonic. 
This contrasts the results obtained for the AT model in the previous sections and the ones obtained for the XX and Ising 
spin-chains \cite{fc-10}.

\begin{figure}[t]
\begin{center}
\includegraphics[width=.7\textwidth]{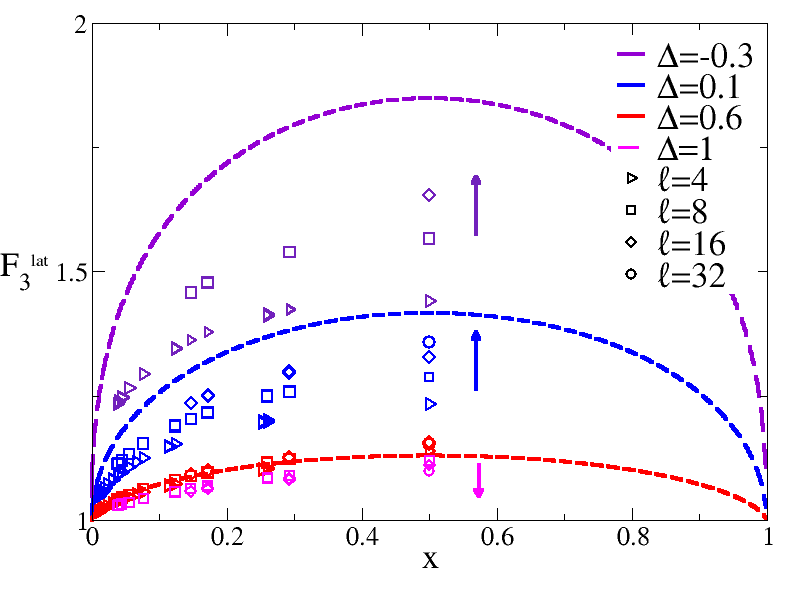}
\end{center}
\caption{TTN data for $F_3^{\rm lat}(x)$ as function of $x$ for various sizes of the chain,
  $\Delta=-0.3,0.1,0.6,1$, and subsystem lengths $\ell=4,8,16,32$. We
  denote with different symbols the values of $\ell$ and with different
  colors the various $\Delta$. The dashed curves are the theoretical
  results given by Eq. (\ref{Fnv}). The arrows denote the
  (asymptotically) increasing subsystem sizes $\ell$.}   
\label{XXZtr_3}
\end{figure}

In order to shed some light on this unexpected phenomenon, it is worth to look at $F_2^{\rm lat} (x)$ as functions of $\ell$ for fixed 
values of $x$.
In Fig.  \ref{XXZfits} we report one of these plots for $x=1/2$. Analogous figures are obtained for other values of $x$.
Corrections to the scaling are non-monotonic in the range $0.2\leq\Delta\leq0.7$.
This phenomenon can be understood if further corrections to the scaling are taken into account.
There are two corrections that can be responsible of this behavior. 
On the one hand,  corrections of the form $\ell^{-m K_L}$ (from $\ell^{-2mK_L/n}$ at $n=2$) for any integer $m$ are know to 
be present \cite{ce-10}, 
on the other hand usual analytic corrections such as $\ell^{-1}$ are generically expected to exist for any quantity from general 
scaling arguments. 
Thus  the most general finite-$\ell$ ansatz has the form
\be
F_2^{\rm lat}(x)=F_2^{\rm CFT}(x)+\frac{f_2(x)}{\ell^{K_L}} +\frac{f_A(x)}{\ell}+\frac{f_B(x)}{\ell^{2K_L}}\dots\,,
\label{ansatz2}
\ee
where  the first correction is the {\it unusual} one employed also  for the Ashkin-Teller model, 
and the other two are the ones just discussed.
The effect of subleading corrections is enhanced by the fact the the amplitude functions $f_2(x)$ and 
$f_A(x)$ or $f_B(x)$ have opposite signs determining the non-monotonic behavior. 
Unfortunately, for values of $\Delta$ for which the effect of subleading corrections is more pronounced (i.e. $0.1\leq\Delta\leq0.6$),
we have $K_L<1<2K_L$, making difficult to disentangle corrections with close exponents. 
Thus, in order to present analyses of a good quality, we ignore the last correction (i.e. we fix $f_B(x)=0$).
To check the proposed scenario, we performed the fit of the data in Fig. \ref{XXZfits} with the ansatz (\ref{ansatz2}) and $f_B(x)=0$.
The results of the fits are reported in the same figure, showing perfect agreement with the data for all the values of $\Delta$.
We repeated the same analysis for other values of $x$, finding the same quality of fits as for $x=1/2$.
However,  we cannot exclude that corrections of the form $\ell^{-2K_L}$ have an important role.

\subsection{Double interval: the $n=3$ case.}

Now we report the same analysis  performed for $\Tr \rho_A^2$ for the third moment of $\rho_A$, i.e. $\Tr \rho_A^3$. 
Again we consider  finite-size XXZ spin-chains and extract the universal function $F^{\rm CFT}_3(x)$ by finite-size analysis. 
The expected CFT result is given for general $n$ by Eq. (\ref{Fnv}). 
In Fig.~\ref{XXZtr_3} we show TTN data for $F^{\rm lat}_3(x)$ (obtained from Eq. (\ref{Flat})) 
at $\Delta=-0.3,0.1,0.6,1$ and subsystem sizes up to $\ell=32$. 
We also show the theoretical curves given by Eq. (\ref{Fnv}). 
As  for the $n=2$, the asymptotic universal curve is approached from below for 
$\Delta\le 0.6$, and from above for $\Delta\geq 0.6$. 
Furthermore, the behavior of the numerical data for $\Delta>0.6$ is non monotonic.
This suggests that the ansatz in Eq. (\ref{ansatz}) is not enough to describe accurately the TTN data and further corrections
to the scaling should be included as for $\Tr \rho_A^2$.

\begin{figure}[t]
\begin{center}
\includegraphics[width=.7\textwidth]{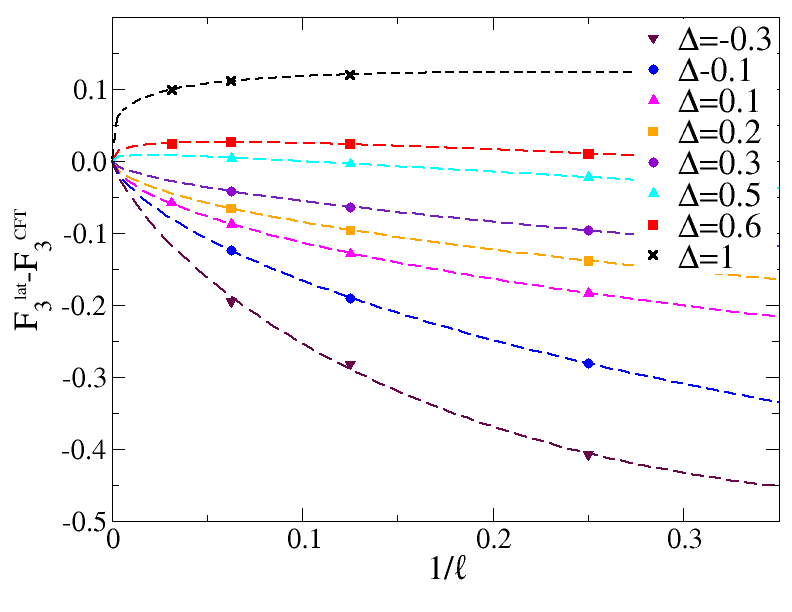}
\end{center}
\caption{TTN data for $F^{\rm lat}_3(x)$ at fixed $x=1/2$ as function of $\ell^{-1}$ for $\ell=4,8,16,32$,
The considered values of $\Delta$ are $\Delta=-0.3,-0.1,01,0.6,1$.  The dashed curves are fits with the ansatz (\ref{f3fits})}  
\label{XXZtr_3x05}
\end{figure}

For $n=3$, the leading corrections to the scaling are described by the ansatz (\ref{ansatz}), i.e. the leading exponent is $2K_L/3$.
Thus, for the cases when subleading corrections are more important (i.e. for $\Delta\geq0.6$) the ordering of the exponents 
is $2K_L/3<4K_L/3<1$ and so it is reasonable to ignore the analytic correction. 
Thus we fit TTN data with the function 
\begin{equation}
F^{\rm lat}_3(x)-F_3^{\rm CFT}(x)=f_3(x)\ell^{-2K_L/3}+f_B(x)\ell^{-4K_L/3}\,. 
\label{f3fits}
\end{equation}
In Fig.~\ref{XXZtr_3x05} we report TTN data for
$F^{\rm lat}_3(x)-F_3^{\textrm{CFT}}(x)$ for $x=1/2$ and several values of $\Delta$. 
The dashed lines are fits with the finite-size ansatz (\ref{f3fits}), that perfectly reproduce the data.

\subsection{Double interval: The von Neumann entropy.}

\begin{figure}[t]
\begin{center}
\includegraphics[width=.8\textwidth]{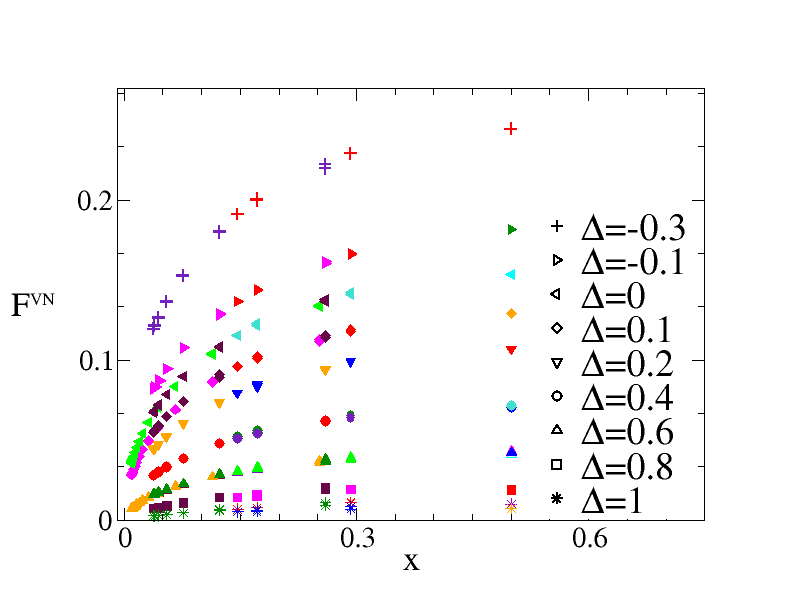}
\end{center}
\caption{TTN data for the von Neumann entropy for  various values of
  $\Delta$ in the interval $[-0.3,1]$. We show with different symbols
  the values of $\Delta$ while different colors stand for different
  $\ell$ and lattice sizes.} 
\label{XXZvn}
\end{figure}

TTN gives access to the full spectrum of the reduced density matrix of $A_1\cup A_2$ and so to the entanglement 
entropy $S_1^{(n)}$ as well. 
In Fig. \ref{XXZvn} we report the function $F_{VN}^{\rm lat}(x)$ defined as
\be
F_{VN}^{\rm lat}(x)=S_{A_1\cup A_2}^{(1)}-S_{A_1}^{(1)}-S_{A_2}^{(1)}-\frac13\log (1-x)\,,
\ee
for $\Delta$ in the interval $[-0.3,1]$ for various $L$ up to 128 and subsystem sizes $\ell=2,4,8,16,32,64$. 
We indicate with different symbols different values of $\Delta$, while the colors are for various sizes $\ell$. 
As known from many other investigations on single and double intervals  (quantum Ising spin chain, XY
model, XXZ) the von Neumann entropy does not show oscillations with the parity of the subsystem and 
the corrections are much smaller, actually negligible from any practical porpouse. 
Fig. \ref{XXZvn} confirms this observation for the two interval entanglement entropy for the XXZ spin-chain in a wide range of $\Delta$.
Indeed, at fixed value of $\Delta$ perfect data collapse is observed even for very small values of $\ell$. 

\begin{figure}[t]
\begin{center}
\includegraphics[width=.8\textwidth]{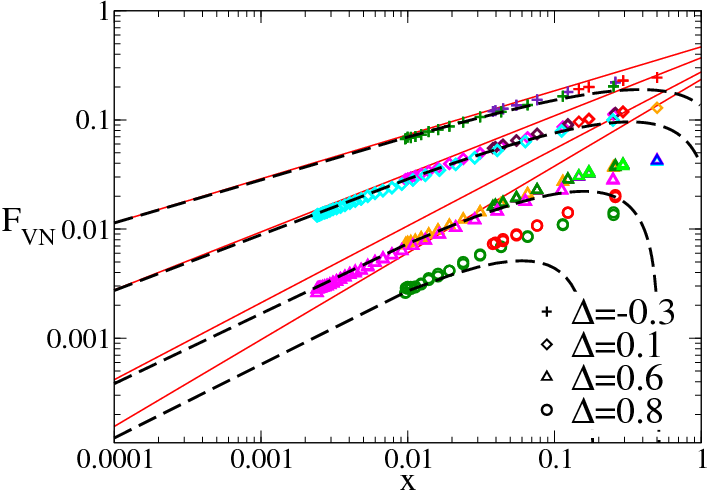}
\end{center}
\caption{TTN data for the von Neumann entropy for
  $\Delta=-0.3,0.1,0.6,0.8$ (the data for different $\Delta$ are denoted
  with different symbols) in log-log scale. We used different colors to
  indicate the different block sizes $\ell$ and lattice sizes $L$. 
  The continuous lines are the small $x$ behavior  obtained from  (\ref{smallx}).
  The dashed lines are the small $x$ behavior where the $O(x)$ term has been added as in Eq. (\ref{smallx2}).}
\label{XXZvn_log}
\end{figure}

Unfortunately, as already stated in the introduction, the CFT prediction for $F_{VN}(x)$ is unknown for general $x$ because 
the analytic continuation of $F_n(x)$ to non-integer $n$ is not achievable. 
However, an expression for the leading term of the small $x$ expansion of $F_{VN}(x)$ has been recently extracted \cite{cct-11}
from Eq. (\ref{s2cft})
\be
F_{VN}(x)=\bigg(\frac{x}{4}\bigg)^\a\sqrt{\pi}\frac{\Gamma(\a+1)}{2\Gamma(\a+3/2)}\,,
\label{smallx}
\ee
where $\a=\textrm{min}[\eta,1/\eta]$ and $\Gamma$ is the Euler function (not to be confused with the $\Gamma$ 
matrix in Eq. (\ref{Gammadef})). 
In order to check the correctness of this formula, in Fig.~\ref{XXZvn_log} we report the same data for $F_{VN}(x)$
in a log-log scale to highlight the power-law behavior for small $x$. 
We also report the  small $x$ expected from Eq. (\ref{smallx}). 
For $\Delta=-0.3$ the agreement is good, but it gets worse increasing $\Delta$. 
The natural explanation is that the considered values of $x$ are not small enough
for the asymptotic Eq. (\ref{smallx}) to be valid. 
We should then include further terms in the small $x$ expansion. 
As explained in Ref. \cite{cct-11}, further coefficients in the expansion for small $x$ are difficult to obtain in general.
However, there is a term that is very easy to obtain and that (luckily enough) is  responsible of the previous disagreement.
Indeed, as shown in Ref. \cite{cct-11} (cf. Eq. 70 and 71 there) the function $F_n(x)$ has always (i.e. independently of $\eta$)
a simple $O(x)$ contribution coming from the denominator in Eq. (\ref{Fnv}), i.e. $|\Theta(0|\Gamma)|^2=1+ x (n-1/n)/6$, that can be
 easily analytically continued giving
\be
F_{VN}(x)=\bigg(\frac{x}{4}\bigg)^\a\sqrt{\pi}\frac{\Gamma(\a+1)}{2\Gamma(\a+3/2)}-\frac{x}3+O(x^{2\a})\,.
\label{smallx2}
\ee
Notice that the added term becomes more important when $\a$ is close to $1$, i.e. in the XXZ spin-chain when $\Delta$
approaches $1$.
In Fig. \ref{XXZvn_log} we also report the prediction (\ref{smallx2}) as dashed line, that is asymptotically in perfect agreement with the 
numerical data for all values of $\Delta$.

\section{Conclusions}
\label{concl}

In this manuscript we provided a number of  results for the asymptotic scaling of the R\'enyi entanglement entropies in
strongly interacting lattice models described by CFTs with $c=1$.
Schematically our results can be summarized as follows.
\begin{itemize}
\item We provided the analytic CFT result for the scaling function $F_2(x)$ for $S_A^{(2)}$ in the case of a  free boson compactified 
on an orbifold describing, among the other things, the scaling limit of the Ashkin-Teller model on the self-dual line.
The final result is given in Eq. (\ref{atf2}).
\item We developed a cluster Monte Carlo algorithm for the two-dimensional Ashkin-Teller model (generalizing the 
procedure of Caraglio and Gliozzi \cite{cg-08} for the Ising model) that gives the scaling functions of the R\'enyi entanglement 
entropy (for integer $n$) of the corresponding one-dimensional quantum model. 
With this algorithm, we calculated numerically the scaling function $F_2(x)$ of the AT model along the self-dual line
and we confirm the validity of the CFT prediction. 
In order to obtain a quantitative agreement, 
the corrections to scaling induced by the finite length of the blocks are properly taken into account.
\item We considered the XXZ spin chains by means of a tree tensor network (TTN) algorithm.
The low-energy excitations of model are described by a free boson compactified on a circle for which CFT predictions are already
available both for $n=2$ \cite{fps-08} and for general integer $n$ \cite{cct-09}.
Taking into account the corrections to the scaling, we confirm these predictions (that resisted until now to
quantitative tests) for $n=2,3$.
Furthermore, we provide numerical determinations of the scaling function of the von Neumann entropy (cf. Fig. \ref{XXZvn})
for which CFT predictions do not exist yet for general $x$. For small $x$ we confirm the recent prediction of Ref. \cite{cct-11} (cf. 
Fig. \ref{XXZvn_log}). 
\end{itemize}

The methods we employed (classical Monte Carlo with cluster observables and TTN) are very general techniques that can be 
easily adapted to other models  of physical interest. On the CFT side, it must be mentioned that a closed form for the functions 
$F_n(x)$ at integer $n$ for a free boson compactified on an orbifold is not yet available, 
but work in this direction is in progress \cite{ct-prep}.

\section*{Acknowledgments}

We thank John Cardy, Maurizio Fagotti, Erik Tonni, Ettore Vicari, and Guifre Vidal for useful discussions.
This work has been partly done when PC was guest of  the Galileo Galilei Institute in Florence whose hospitality is kindly 
acknowledged.

\end{document}